\documentclass[preprint,12pt]{elsarticle}
\usepackage{amsmath,amssymb,amsfonts}
\usepackage{algorithmic}
\usepackage{graphicx}
\usepackage{textcomp}

\usepackage{algorithm}
\usepackage{array}
\usepackage{stfloats}
\usepackage{url}
\usepackage{verbatim}
\usepackage{color}
\usepackage{microtype}
\usepackage{booktabs}
\usepackage{hyperref}
\usepackage{mathtools,mathrsfs}
\usepackage{multirow}

\journal{CMPB}

\begin{document}

\begin{frontmatter}



\title{Learn Single-horizon Disease Evolution for Predictive Generation of  Post-therapeutic Neovascular Age-related Macular Degeneration\tnoteref{label1}}


\author[a1]{Yuhan Zhang}
\ead{zhangyuhan@njust.edu.cn}

\author[a1]{Kun Huang}
\ead{huangkun@njust.edu.cn}

\author[a1]{Mingchao Li}
\ead{chaosli@njust.edu.cn}

\author[a2]{Songtao Yuan}
\ead{yuansongtao@vip.sina.com}

\author[a1]{Qiang Chen\corref{cor5}}
\ead{chen2qiang@njust.edu.cn}
\cortext[cor5]{Qiang Chen is the corresponding author of this work.}

\affiliation[a1]{organization={School of Computer Science and Engineering},
            addressline={Nanjing University of Science and Technology}, 
            city={Nanjing},
            postcode={210094}, 
            country={China}}

\affiliation[a2]{organization={Department of Ophthalmology},
	addressline={The First Affiliated Hospital with Nanjing Medical University}, 
	city={Nanjing},
	postcode={210094}, 
	country={China}}

\begin{abstract}
\textbf{Background and Objective:}
Most of the existing disease prediction methods in the field of medical image processing fall into two classes, namely image-to-category predictions and image-to-parameter predictions. 
Few works have focused on image-to-image predictions.
Different from multi-horizon predictions in other fields, ophthalmologists prefer to show more confidence in single-horizon predictions due to the low tolerance of predictive risk. 

\textbf{Methods:}
We propose a single-horizon disease evolution network (SHENet) to predictively generate post-therapeutic SD-OCT images by inputting pre-therapeutic SD-OCT images with neovascular age-related macular degeneration (nAMD). 
In SHENet, a feature encoder converts the input SD-OCT images to deep features, then a graph evolution module predicts the process of disease evolution in high-dimensional latent space and outputs the predicted deep features, and lastly, feature decoder recovers the predicted deep features to SD-OCT images. 
We further propose an evolution reinforcement module to ensure the effectiveness of disease evolution learning and obtain realistic SD-OCT images by adversarial training.

\textbf{Results:}
SHENet is validated on 383 SD-OCT cubes of 22 nAMD patients based on three well-designed schemes (P-0, P-1 and P-M) based on the quantitative and qualitative evaluations.
Three metrics (PSNR, SSIM, 1-LPIPS) are used here for quantitative evaluations.
Compared with other generative methods, the generative SD-OCT images of SHENet have the highest image quality (P-0: 23.659, P-1: 23.875, P-M: 24.198) by PSNR.
Besides, SHENet achieves the best structure protection (P-0: 0.326, P-1: 0.337, P-M: 0.349) by SSIM and content prediction (P-0: 0.609, P-1: 0.626, P-M: 0.642) by 1-LPIPS.
Qualitative evaluations also demonstrate that SHENet has a better visual effect than other methods.

\textbf{Conclusions:} 
SHENet can generate post-therapeutic SD-OCT images with both high prediction performance and good image quality, which has great potential to help ophthalmologists forecast the therapeutic effect of nAMD.
\end{abstract}


%%\item Research highlight 1
%%\item Research highlight 2


\end{frontmatter}


\section{Introduction}

\subsection{Application Background}
Neovascular age-related macular degeneration (nAMD) is a main subtype of AMD. 
As the intravitreal vascular endothelial growth factor (VEGF) level elevates, choroidal neovasculars invade the avascular outer retina and severely damage photoreceptors, resulting in rapid vision loss \cite{01foss2022development}. 
At present, anti-VEGF injection is considered the preferred nAMD therapy \cite{02tadayoni2021brolucizumab}. 
Although ophthalmologists always give anti-VEGF injections after nAMD diagnosis, nAMD does not always respond satisfactorily to treatment. 
Besides, due to the lack of uniform guidelines, it is difficult for ophthalmologists to predict the short-term therapeutic response after anti-VEGF injection according to their subjective experiences \cite{03mettu2021incomplete}. 
This results in huge economic pressures and waste of resources \cite{04maguire2016five}. 
Therefore, based on the known pre-therapeutic status of nAMD at time point $t_1$, predicting the post-therapeutic status of nAMD at time point $t_2 = t_1 + \Delta t$ can effectively forecast the efficacy of anti-VEGF injection for each patient. 
This can promote better clinical decision making.

\subsection{SD-OCT Imaging Brief}
\begin{figure}[!t]
	\centering
	\includegraphics[width=1.0\textwidth]{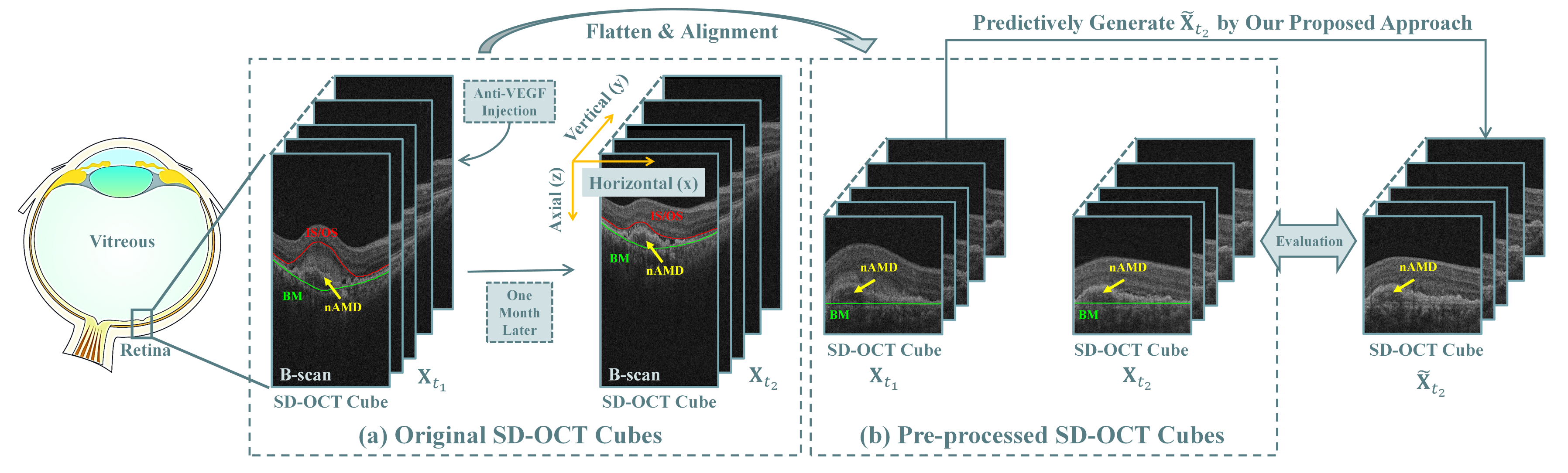}
	\caption{\textbf{Clinical Application of Our Proposed Approach.} (a) shows the workflow of neovascular age-related macular degeneration (nAMD) therapy. Ophthalmologists diagnose nAMD by SD-OCT imaging at time point $t_1$ and then give anti-VEGF injection for treatment. Generally one month later ($t_2 = t_1 + \Delta t$), ophthalmologists can estimate the therapy response by SD-OCT imaging again. Red line and green line indicate two retinal layer surfaces, which will be used for data pre-processing (flatten and alignment). (b) shows the clinical application of our proposed approach. Based on pre-processed SD-OCT images at time point $t_1$, our proposed approach predictively generates SD-OCT images at time point $t_2$ to help ophthalmologists forecast the therapy response of anti-VEGF injection, and consequently make more reasonable treatment decisions.}
	\label{fig:1}    % Give a unique label
\end{figure}

Spectral-domain optical coherence tomography (SD-OCT) is a noninvasive, depth-resolved, high-resolution, and volumetric imaging technique. 
SD-OCT has become a pivotal diagnostic tool to visualize and quantitatively evaluate retinal morphological changes \cite{05lan2021design}, including the diagnosis and tracing of nAMD \cite{07gharbiya2018comparison,08saito2017efficacy}. 
Each SD-OCT imaging can produce 3D volumetric images, also known as a \textbf{cube}.
As shown in Fig. \ref{fig:1}(a), we take Cirrus SD-OCT device as the example, each SD-OCT cube contains $1024 \times 512 \times 128$ voxels with a corresponding trim size of $2mm \times 6mm \times 6mm$ on the retina. 
In a cube, each slice along the vertical direction, with the size of $1024 \times 512$, is known as a \textbf{B-scan}. 
A complete SD-OCT cube contains 128 continuous B-scans in space.

\subsection{Dilemma of Multi-horizon Predictions on Medical Images}
\begin{figure}[!t]
	\centerline{\includegraphics[width=0.8\columnwidth]{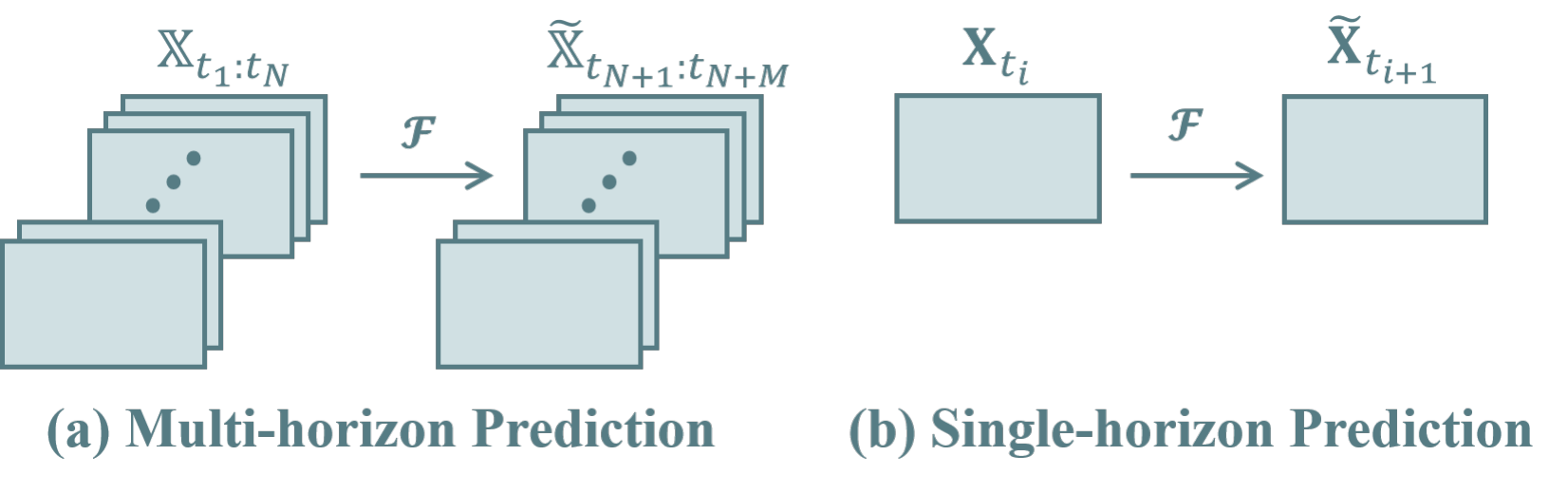}}
	\caption{\textbf{Prediction Visualization.} (a) is multi-horizon prediction, (b) is single-horizon prediction.}
	\label{fig:5}    % Give a unique label
\end{figure}

Diseases-associated predictions are more restrictive than general predictions. 
In the field of pattern recognition, multi-horizon predictions have been widely applied for natural language predictions, action predictions, video predictions, traffic predictions and etc. 
Given $\mathbb{X}_{{t_1}:{t_N}}=[{\bf X}_{t_1},{\bf X}_{t_2}, \cdots, {\bf X}_{t_N}] \in {\mathbb{R}}$ as the historical \textit{N} observations, each observation is obtained at the different time point and the time interval between any two adjacent time point is uniform. 
The actual future \textit{M} observations are formally expressed as $\mathbb{X}_{{t_{N+1}}:{t_{N+M}}}=[{\bf X}_{t_{N+1}},{\bf X}_{t_{N+2}}, \cdots, {\bf X}_{t_{N+M}}] \in {\mathbb{R}}$. 
We expect multi-horizon predictions to learn a mapping function $\mathcal{F}: \mathbb{X}_{{t_1}:{t_N}} \rightarrow \widetilde{\mathbb{X}}_{{t_{N+1}}:{t_{N+M}}}$ to obtain the prediction result $\widetilde{\mathbb{X}}_{{t_{N+1}}:{t_{N+M}}}$ as close as $\mathbb{X}_{{t_{N+1}}:{t_{N+M}}}$, as shown in Fig. \ref{fig:5}(a). 
However, when applying multi-horizon predictions on medical images, several actual challenges raise:

\begin{itemize}
	\item
	Obtaining long-series observations from the same patient is intractable in practical clinical scenes, and it also is necessary to make effective predictions for a new patient with only one observation at the current time point.
	\item 
	For serial medical data, given any two different time points ${t_i},{t_j} \in ({t_2},{t_N}]$, time intervals from adjacent time points may be different, namely ${t_i}-{t_{i-1}} \not\equiv {t_j}-{t_{j-1}}$.
	\item
	Most medical observations generate 3D data and it is expensive for GPUs to learn from serial 3D data.
	\item
	Therapeutic intervention dependent on medicine injection at a random time point is a key factor that cannot be ignored for diseases-associated predictions, and most medicine injection treatments only work for a short period of time.
	\item
	It is well-known that the prediction accuracy decreases over time and risk tolerance on medical predictions is lower than other predictions, so clinicians always pay more confidence on the short horizon than the longer horizon.
\end{itemize}

These difficulties make it difficult for multi-horizon predictions to be really applied to medical images, thus learning a single-horizon prediction for medical images is more practical and realistic.
Given one historical observation ${\bf X}_{t_i} \in {\mathbb{R}}$ at time point $t_i$, single-horizon prediction learns a mapping function $\mathcal{F}: {\bf X}_{t_i} \rightarrow \widetilde{{\bf X}}_{t_{i+1}}$ to obtain the prediction result at time point $t_{i+1}$, as shown in Fig. \ref{fig:5}(b).

\subsection{Target of Our Work}
Nowadays, most existing disease prediction methods in the field of medical image processing fall into two classes, namely classification-based image-to-category (I2C) predictions, and regression-based image-to-parameter (I2P) predictions. 
Few works have focused on generation-based image-to-image (I2I) predictions, even though generative adversarial networks (GANs) have been widely used for modality transformation between medical images from different imaging devices.
The aim of post-therapeutic prediction is to generate an image that explains how the anatomical appearance changes after treatment. 
We consider that the predictive post-therapeutic SD-OCT images would enable a better understanding of nAMD and clinical decision-making by presenting a visual post-therapeutic status of nAMD.
According to the above, the target of our work is defined as:

Given a serial SD-OCT cubes with fixed time interval $\Delta t$, writing as $\mathbb{X}=[{\bf X}_{t_1},{\bf X}_{t_2}, \cdots ,{\bf X}_{t_{N+M}}] \in {\mathbb{R}}$, and any time point \textit{$t_i$} can be represented as ${t_i} = {t_1} + (i-1) \Delta t$, $i \in [1,N+M]$. 
Anti-VEGF injection is given at each time point. 
For any SD-OCT cube ${\bf X}_{t_i}$, we hope to learn a mapping function $\mathcal{F}:{\bf X}_{t_i} \rightarrow \widetilde{\bf{X}}_{t_{i+1}}$ to predictively generate $\widetilde{\bf{X}}_{t_{i+1}}$ as close as ${\bf{X}}_{t_{i+1}}$.

However, learning the mapping function $\mathcal{F}$ directly based on a 3D SD-OCT cube is difficult and cost-consuming. 
Thus, for the SD-OCT cube ${\bf X}_{\it t_i} = [{\boldsymbol x}_{t_{i}}^{1}, {\boldsymbol x}_{t_{i}}^{2}, \cdots, {\boldsymbol x}_{t_{i}}^{128}]$, where $\boldsymbol{x}_{t_i}^j$ is \textit{j}-th B-scan in ${\bf X}_{t_{i}}$, we learn the 2D mapping function $\mathcal{F}_{2D}:{\boldsymbol x}_{t_i}^j \rightarrow \widetilde{\boldsymbol{x}}_{{t}_{i+1}}^j$. 
$\mathcal{F}_{2D}$ is carried out 128 times until the SD-OCT cube is predicted completely.

\subsection{Contributions}
In this paper, we present a \textbf{S}ingle-\textbf{H}orizon disease \textbf{E}volution \textbf{Net}work (SHENet) to predictively generate the post-therapeutic SD-OCT images based on pre-therapeutic SD-OCT images with nAMD. 
SHENet can help forecast the short-term response of anti-VEGF injection for individual nAMD patients. 
The main contributions in this paper are summarized as: 
\begin{itemize}
	\item 
	We explore the possibility of predictively generating post-therapeutic SD-OCT images based on pre-therapeutic SD-OCT images with nAMD, and further propose SHENet to solve this problem.
	\item
	Graph evolution module (GEM) is proposed to imprison the process of disease evolution in the high-dimensional latent space by graph representation learning.
	\item
	Evaluation reinforcement module (ERM) is proposed to reinforce the disease evolution process by combining an additional reconstruction generator and contrastive learning.
	\item
	We design targeted experimental schemes according to actual clinical realities, and the results demonstrate that SHENet has great potential to generate post-therapeutic SD-OCT images with both high prediction performance and good image quality.
\end{itemize}

%---------------------------------------------------------------------
\section{Related Work}

\subsection{Predictions on AMD}
In clinical scenarios, several prediction requirements have been raised around AMD by ophthalmologists. In present, there are no uniform guidelines to be helpful for making predictions and ophthalmologists rely only on their own rich clinical experiences. Thus, ophthalmologists warrant the need for objective outcomes of subjective predictions. AMD typically develops from an early to an advanced form and advanced AMD is difficult to be cured effectively. When AMD is in its early stage, ophthalmologists predict the risk of progression to advanced AMD within the future short term in order to adapt therapies, recommendations, and follow-up frequency  \cite{09yim2020predicting,10ajana2021predicting,11banerjee2020prediction,12yan2021genome,13&17bhuiyan2020artificial}. For advanced nAMD, ophthalmologists predict best-corrected visual acuity (BCVA) outcomes in patients receiving standard therapy  \cite{14schmidt2018machine,15rohm2018predicting,16diack2021baseline}. For geographic atrophy (GA), as non-neovascular advanced AMD, there is a lack of effective treatments for retinal areas that have progressed to GA. But predicting the GA progression \cite{18rossant2021normalization,19zhang2021integrated,20reiter2020investigating,21nattagh2020oct,22zhang2019multi,23yang2021multi} could allow for a better understanding of the pathogenesis and forewarn preventive treatment to normal retinal areas that are at high risk of developing GA in the future. Besides, several other predictions also reveal clinical needs. Bogunovic et al. \cite{24bogunovic2017prediction} predicted low and high anti-VEGF injection requirements based on sets of SD-OCT images acquired during the initiation phase in nAMD. Liu et al. \cite{25liu2020prediction} and Lee et al. \cite{47lee2021post} exploringly generated individualized post-therapeutic SD-OCT images that could predict the short-term response of anti-VEGF injection for nAMD based on pre-therapeutic images using Pix2Pix. Forshaw et al. \cite{26forshaw2020full} predicted the visual gain from cataract surgery when the main cause of vision loss is nAMD. 
Pham et al. \cite{48pham2022generating} generated future fundus images for early age-related macular degeneration based on generative adversarial networks.

\subsection{Generative Adversarial Networks (GANs)}
Generative adversarial networks (GANs) \cite{27goodfellow2014generative} have become one of the widely leveraged techniques for generating images that look like the real thing. 
Traditional L1 or L2 supervision often results in blurred images \cite{28johnson2016perceptual}, but GANs introduce an additional discriminator to play a min-max game with the generator that enforces the generator to output more realistic images. 
The discriminator is mainly used to estimate the divergence difference between generated fake images and real images. Different adversarial losses estimate different divergences between Wasserstein divergence distribution \cite{29arjovsky2017wasserstein} and f-divergence family distribution \cite{30nowozin2016f}. 
If target domains own multiple distributions, conditional GAN (cGAN) \cite{31mirza2014conditional} uses conditional labels to guide the generator to produce the images by fitting the specified distribution. 
For example, Yoo et al. \cite{51yoo2020generative} proposed a postoperative appearance prediction model for orbital decompression surgery for thyroid ophthalmopathy using a conditional GAN.
GANs are also used for image-to-image (I2I) translation, in which pixel-level losses or feature-level losses are embedded to ensure the quality and stability of the generated image \cite{48pham2022generating,49qiu2022improved,50zhang2022bpgan}. 
Besides, literatures \cite{35schonfeld2020u, 36park2019semantic} explored the impact of discriminator architecture on GANs.

%---------------------------------------------------------------------
\section{Methods}

This study was approved by the ethic committee of the First Affiliated Hospital with Nanjing Medical University.

\subsection{Pre-processing: Voxel-wise Serial Alignment}
The random deviation of rotation angle and displacement among SD-OCT cubes at different time points caused by human operations cannot be learned, which further limits the I2I prediction. 
In other words, for two original SD-OCT cubes from the same patient captured at different time points, the B-scan with the same index in two cubes may not correspond to the same anatomical tissue. 
Thus, we perform voxel-wise serial alignment, including image flattening in vertical direction and image alignment in horizontal-axial directions, for all SD-OCT cubes before running SHENet. 
In this way, all SD-OCT cubes at different time points are aligned in the 3D space. 
The aligned SD-OCT cube can be seen in Fig. \ref{fig:1}(b).

\textbf{Image Flattening in Vertical Direction.} 
We first obtain the locations of Bruch’s membrance (BM) of all B-scans by layer segmentation approach \cite{37zhang2021robust}, and then all B-scans are flattened based on BM. 
Finally, we crop the B-scan restricted to the region from 0.75mm above BM to 0.25mm below BM. 
In this processing, negligible vitreum regions and sclera regions are removed as much as possible. 
Image flattening also reduces the size of the input SD-OCT images to alleviate the memory pressure of hardware.

\textbf{Image Alignment on Horizontal-Axial Directions.}
We first generate the 2D vessel fundus images of each SD-OCT cube by restricting the projection region between inner segments/outer segments (IS/OS) and BM.
The vessel fundus images are aligned using a scale-invariant feature transform (SIFT) flow method to obtain the transformation matrixes.
Lastly, the transformation matrixes are applied to horizontal-axial directions of SD-OCT cubes to obtain the aligned SD-OCT cubes.

\begin{figure}[t]
	\centering
	\includegraphics[width=1.0\textwidth]{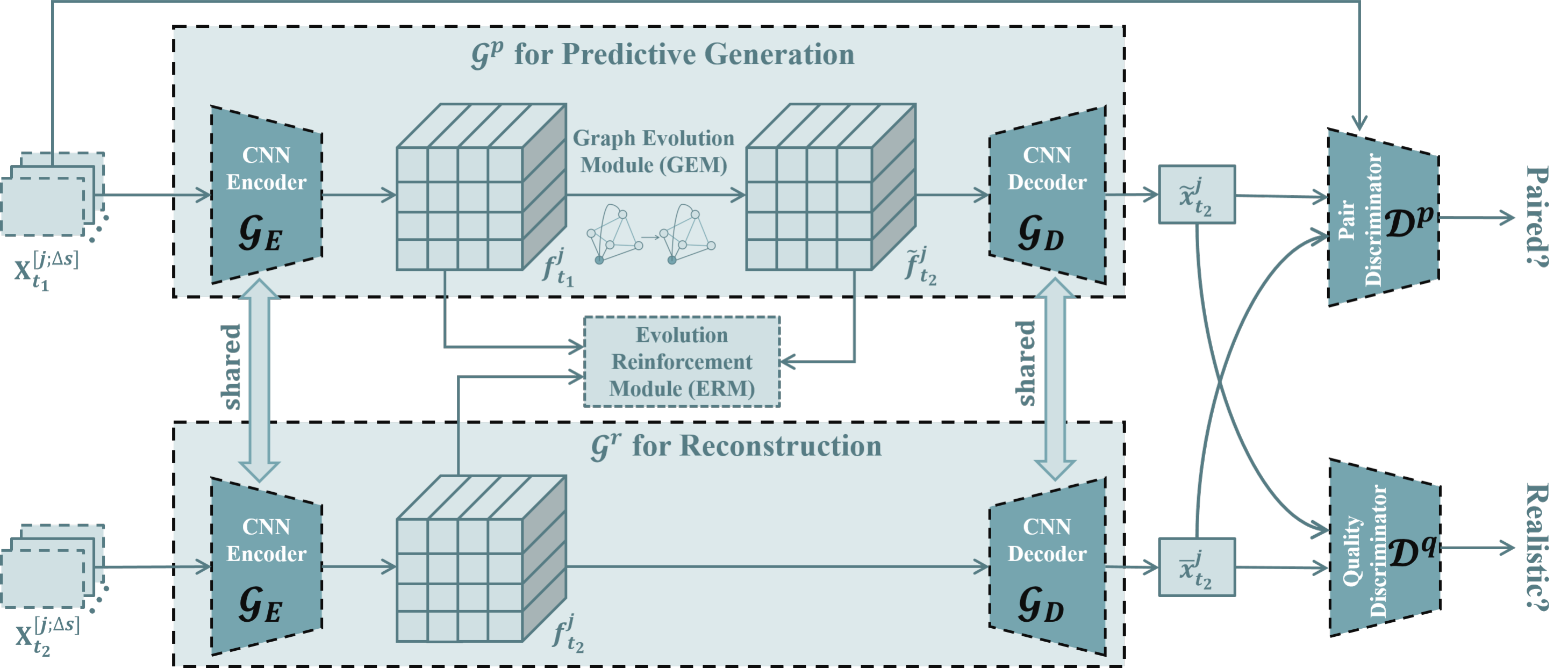}
	\caption{\textbf{Overview of SHENet in Training Stage.} SHENet consists of a prediction generator $\mathcal{G}^p$, a reconstruction generator $\mathcal{G}^r$, an evolution reinforcement module (ERM), a quality discriminator $\mathcal{D}^q$ and a pair discriminator $\mathcal{D}^p$. A little piece of B-scans $\mathbf{X}_{t_{1}}^{[j ; \Delta s]} \in \mathbf{X}_{t_{1}}$ as multi-channel inputs are sent to $\mathcal{G}^p$ to predictively generate $\widetilde{\boldsymbol{x}}_{t_2}^j$, meanwhile a little piece of B-scans $\mathbf{X}_{t_{2}}^{[j ; \Delta s]} \in \mathbf{X}_{t_{2}}$ as multi-channel inputs are sent to $\mathcal{G}^r$ to reconstruct $\overline{\boldsymbol{x}}_{t_2}^j$. ERM reinforces the disease evolution in the high-dimensional latent space. $\mathcal{D}^q$ ensures the image quality of $\widetilde{\boldsymbol{x}}_{t_2}^j$ and $\overline{\boldsymbol{x}}_{t_2}^j$ is realistic, and $\mathcal{G}^p$ decides whether $\widetilde{\boldsymbol{x}}_{t_2}^j$ and $\overline{\boldsymbol{x}}_{t_2}^j$ are paired with $\boldsymbol{x}_{t_1}^j$ in terms of pathological characterization that reflects therapy response. In the inference stage of SHENet, only prediction generator $\mathcal{G}^p$ remains and other all components are removed.}
	\label{fig:2}    % Give a unique label
\end{figure}

\subsection{Overview of SHENet}
Pix2Pix \cite{32isola2017image} has been a great success to apply conditional GAN (cGAN) to supervised I2I translation. It regards input images as additional conditions to learn an I2I mapping, consequently producing specified output images. 
SHENet further extends Pix2Pix for our nAMD prediction task and its overview is illustrated in Fig. \ref{fig:2}. 
In terms of model architecture, SHENet consists of:

\textbf{(1) Prediction Generator} $\mathcal{G}^{p}$ is the core of SHENet, including a feature encoder, a graph evolution module (GEM), and a feature decoder, which can predictively generate post-therapeutic SD-OCT images by inputting pre-therapeutic SD-OCT images with nAMD.

\textbf{(2) Reconstruction Generator} $\mathcal{G}^{r}$ is an auxiliary generator in the training process and will be removed in the model inference stage. 
$\mathcal{G}^{r}$ removes GEM from $\mathcal{G}^{p}$ and utilizes a reconstruction task to help $\mathcal{G}^{p}$ imprison the process of disease evolution in the high-dimensional latent space, and further to distill the function of feature encoder and decoder. 

\textbf{(3) Evolution Reinforcement Module (ERM)} reinforces the process of disease evolution by working with $\mathcal{G}^{r}$ based on contrastive learning.

\textbf{(4) Quality Discriminator} $\mathcal{D}^{q}$ ensures that the predicted and reconstructed images look realistic.

\textbf{(5) Pair Discriminator} $\mathcal{D}^{p}$ determines whether predicted and reconstructed images are paired with input images in terms of pathological characterization that reflects therapy response.

\subsubsection{\textbf{Motivation: Why learn the disease evolution in the high-dimensional latent space?}}
Liu et al. \cite{25liu2020prediction} and Lee et al. \cite{47lee2021post} have preliminarily used Pix2Pix to generate post-therapeutic SD-OCT images and show the feasibility, but learning a pixel-to-pixel prediction on SD-OCT images seems not to be a delicate work. 
Severe speckle noise in SD-OCT images and the imbalanced proportion of foreground pixels (\textit{i.e.} nAMD) relative to the background pixels (\textit{i.e.} non-nAMD) degenerate the pixel-to-pixel prediction performance. 
Compared with Pix2Pix, SHENet imprisons the process of disease evolution in the high-dimensional latent space:
\begin{equation}
	\begin{split}
		&Pix2Pix: {\boldsymbol x}_{t_1}^j \stackrel{Generator}{\longrightarrow} \widetilde{\boldsymbol{x}}_{t_2}^j \\
		&SHENet: {\boldsymbol x}_{t_1}^j \stackrel{Enc}{\longrightarrow} 
		{\boldsymbol f}_{t_1}^j 
		\stackrel{Pred}{\longrightarrow} \widetilde{\boldsymbol{f}}_{t_2}^j 
		\stackrel{Dec}{\longrightarrow} 
		\widetilde{\boldsymbol{x}}_{t_2}^j
		\label{eq:1}
	\end{split}
\end{equation}
where ${\boldsymbol f}_{t_1}^j$ is the encoding features of ${\boldsymbol x}_{t_1}^j$ and $\widetilde{\boldsymbol{f}}_{t_2}^j$ is the predicted features of ${\boldsymbol x}_{t_1}^j$. 
The high-dimensional features are more condensed and effective, because the invalid background and noise information are removed, and distinct disease information is retained. 
Therefore, we consider that learning the process of disease evolution after treatment in the high-dimensional latent space is more reasonable than performing a pixel-to-pixel prediction.

\subsubsection{\textbf{Multiple B-scans As Model Input.}}
Given an SD-OCT cube at time point ${t_1}$ as 
$\mathbf{X}_{t_{1}}=[{\boldsymbol{x}_{t_{1}}^{1}, \boldsymbol{x}_{t_{1}}^{2}, \cdots , \boldsymbol{x}_{t_{1}}^{128}]}$
and aligned SD-OCT cube at time point ${t_2} = {t_1} + \Delta t$ as  $\mathbf{X}_{t_{2}}=[{\boldsymbol{x}_{t_{2}}^{1}, \boldsymbol{x}_{t_{2}}^{2}, \cdots , \boldsymbol{x}_{t_{2}}^{128}]}$. In general, we should train a 2D mapping $\mathcal{F}_{2D}$ to predictively generate $\widetilde{\boldsymbol{x}}_{t_2}^j$ as close as real $\boldsymbol{x}_{t_2}^j$:
\begin{equation}
	\mathcal{F}_{2D} : {\boldsymbol x}_{t_1}^j \rightarrow \widetilde{\boldsymbol{x}}_{t_2}^j
	\label{eq:2}
\end{equation}
However, in SD-OCT images, ${\boldsymbol x}_{t_1}^j$ with similar pathological characterization may develop to $\boldsymbol{x}_{t_2}^j$ with different pathological characterization, resulting in difficult model convergence and random prediction results. 
For example, for patient-1, a healthy SD-OCT B-scan at time point $t_1$ evolves to an SD-OCT B-scan with nAMD at time point $t_2$. However, for patient-2, a healthy SD-OCT B-scan at time point $t_1$ may remain its healthy status at time point $t_2$.
Thus, to speed up the model convergence and improve the model robustness, we choose to stack a little piece of B-scans of $\mathbf{X}_{t_{1}}$ as $\mathbf{X}_{t_{1}}^{[j ; \Delta s]}=[\boldsymbol{x}_{t_{1}}^{j - \Delta s} , \cdots , \boldsymbol {x}_{t_{1}}^j , \cdots , \boldsymbol{x}_{t_{1}}^{j + \Delta s}] \in \mathbf{X}_{t_{1}}$ as multi-channel inputs to predict $\widetilde{\boldsymbol{x}}_{t_2}^j$:
\begin{equation}
	\mathcal{F}_{2D} : 
	{\bf X}_{t_1}^{[j ; \Delta s]} 
	\rightarrow 
	\widetilde{\boldsymbol{x}}_{t_2}^j
	\label{eq:3}
\end{equation}
$\mathcal{F}_{2D}$ slides on each B-scan of ${\bf X}_{t_1}$ until $\widetilde{\bf{X}}_{t_2}$ is predicted completely. 

\begin{figure}[!t]
	\centerline{\includegraphics[width=0.85\columnwidth]{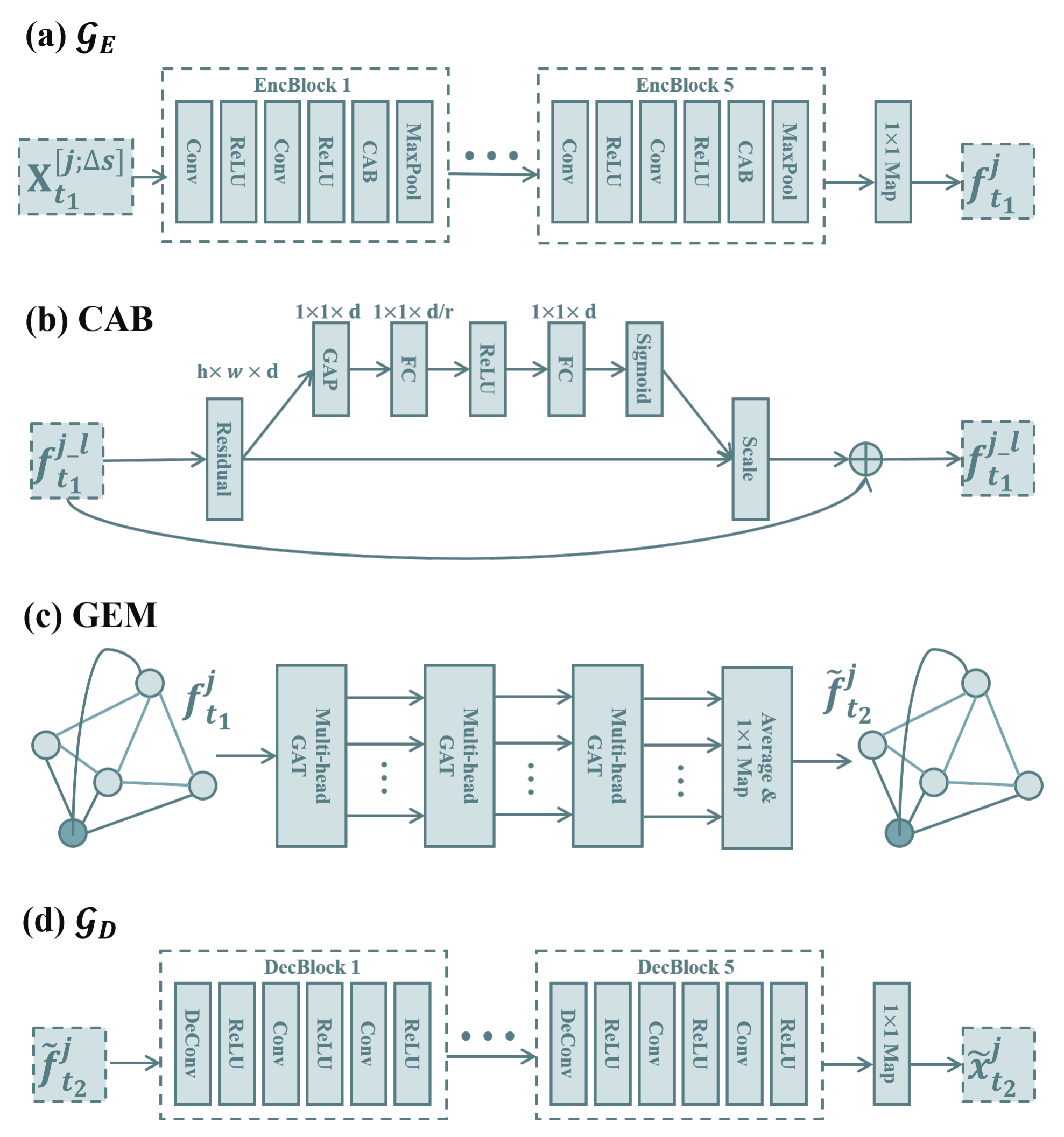}}
	\caption{\textbf{Details of Prediction Generator} $\mathbf{\mathcal{G}^{\it p}}$. (a) is feature encoder $\mathcal{G}_E$ consisting of several customized encoding blocks. (b) is channel attention block (CAB), where GAP indicates global average pooling. (c) is graph evolution module (GEM) that stacks three multi-head GAT modules. (d) is feature decoder $\mathcal{G}_D$ consisting of several customized decoding blocks.}
	\label{fig:3}    % Give a unique label
\end{figure}

\subsection{Prediction Generator $\mathcal{G}^p$ of SHENet}
The prediction generator $\mathcal{G}^p$ of SHENet consists of three modules: feature encoder ($\mathcal{G}_{E}$), graph evolution module (GEM), and feature decoder ($\mathcal{G}_{D}$). 
The pre-therapeutic SD-OCT images are mapped to high-dimensional latent space by $\mathcal{G}_{E}$, then GEM predicts the process of disease evolution after treatment, and finally, $\mathcal{G}_{D}$ recovers the predicted features to post-therapeutic SD-OCT images.

\textbf{Feature Encoder ($\mathcal{G}_{E}$).}
Fig. \ref{fig:3}(a) illustrates the architecture of feature encoder $\mathcal{G}_{E}$ that stacks several encoding blocks (EncBlock) and a mapping layer. 
Each EncBlock consists of two 3$\times$3 convolutional layers with ReLU activation, one channel attention block (CAB) \cite{39hu2018squeeze}, and one max-pooling layer for down-sampling. 
CAB (Fig. \ref{fig:3}(b)) improves the quality of features by explicitly modeling the interdependencies between the channels of its convolutional features. 
Through CAB, informative features are selectively emphasized and less useful ones are suppressed.

\textbf{Graph Evolution Module (GEM).}
Given the encoding features ${\boldsymbol f}_{t_1}^j \in {\mathbb R}^{{h'} \times {w'} \times {d'}}$ by $\mathcal{G}_{E}$, where $P={h'} \times {w'}$ is feature numbers and ${d'}$ is feature dimension. 
We consider the evolution of each feature should be related to other features.
Intuitively, a series of convolution layers or a linear regression is easier to implement to predict the change in the feature level. 
However, a series of convolution layers or a linear regression defaults that all other features contribute equally to the targeted feature. 
In fact, the contributions of all features are unequal due to their different semantic information. 
Graph neural networks can model the contributions of all features automatically, which is more reasonable than a series of convolution layers or a linear regression.
Let each feature be the vertex of the graph, and we can represent ${\boldsymbol f}_{t_1}^j$ as a fully-connected undirected graph $\mathcal{R} = (\mathcal{V},\mathcal{E})$, where $\mathcal{V}$ is vertex set and $\mathcal{E}$ is edge set. 
Further given the adjacency matrix ${\boldsymbol A}$, the diagonal degree matrix ${\boldsymbol D}$ and the identity matrix ${\boldsymbol I}$, the relation of features can be extracted by the following formula:
\begin{equation}
	{\boldsymbol H}^{l+1} = 
	\sigma (\widetilde{\boldsymbol{A}} {\boldsymbol H}^{l} {\boldsymbol W}^{l}) , 
	\widetilde{\boldsymbol{A}} = 
	{\boldsymbol D}^{- \frac{1}{2}} (\boldsymbol{A} + \boldsymbol{I}) {\boldsymbol D}^{- \frac{1}{2}}
	\label{eq:4}
\end{equation}
where ${\boldsymbol W}^{l} \in {\mathbb R}^{{d_{in}} \times {d_{out}}}$ is the learnable weight, and $\sigma$ is the \textit{Mish} function \cite{40misra2019mish}. 
${\boldsymbol H}^{l} \in {\mathbb R}^{{P} \times {d_{in}}}$, ${\boldsymbol H}^{l+1} \in {\mathbb R}^{{P} \times {d_{out}}}$ are the input features and the updated features at \textit{l}-th layer. 
Intuitively, for \textit{p}-th vertex $\mathcal{V}_{p}$, its all neighbors contribute unequally to the evolution of  $\mathcal{V}_{p}$. 
To model this, we introduce the multi-head GAT \cite{41velivckovic2017graph} to explicitly consider the importance of the neighbors. 
Take independent GAT as an example, we calculate the attentive score ${\gamma}_{pq}$ of the vertex pair $(\boldsymbol{h}_{p},\boldsymbol{h}_{q})$:
\begin{equation}
	{\gamma}_{\it pq} = \frac{exp(LReLU(\boldsymbol{\alpha}^T [\boldsymbol{W} \boldsymbol{h}_{p},\boldsymbol{W} \boldsymbol{h}_{q}]))}{\sum_{{k} \in \mathcal{N}_{p}} exp(LReLU(\boldsymbol{\alpha}^T [\boldsymbol{W} \boldsymbol{h}_{p},\boldsymbol{W} \boldsymbol{h}_{k}]))}
	\label{eq:5}
\end{equation}
where $\mathcal{N}_{p}$ is the neighbors of \textit{p}-th vertex in the graph, and $[\cdot]$ is a concatenation operation. 
$\boldsymbol{W} \in {\mathbb R}^{{d}_{out} \times {d}_{in}}$ indicates a linear mapping and $\boldsymbol{\alpha} \in {\mathbb R}^{{2d_{out}}}$ indicates a single-layer fully-connected layer. 
\textit{LReLU} is the nonlinear activation. 
Then, we compute the average of multi-head GAT for the output of vertex $\mathcal{V}_{p}$:
\begin{equation}
	\widetilde{\boldsymbol{h}}_p = \sigma (\frac{1}{G} \sum\limits_{g=1}^{G} \sum\limits_{\sum_{{q} \in \mathcal{N}_{i}}} {\gamma}_{pq}^g \boldsymbol{W}^g \boldsymbol{h}_p)
	\label{eq:6}
\end{equation}
where \textit{G}=5 is the number of heads. 
Similarly, we repeat the GAT computation on each vertex to obtain the complete output. 
Considering the powerful ability of graph neural networks for information inference, GEM is only composed of 3 multi-head GATs with the channel numbers 1024, 1024, 1024. 
The overview of GEM is shown in Fig. \ref{fig:3}(c).

\textbf{Feature Decoder ($\mathcal{G}_{D}$).}
Fig. \ref{fig:3}(d) illustrates the architecture of feature decoder $\mathcal{G}_{D}$ that stacks several decoding blocks (DecBlock) and a mapping layer. 
Each DecBlock consists of one de-convolution layer with ReLU activation, and two convolutional layers with ReLU activation.

\subsection{Evolution Reinforcement Module (ERM)}
To reinforce the process of disease evolution, we introduce reconstruction generator $\bf{\mathcal{G}}^{r}$ that removes GEM from $\bf{\mathcal{G}}^{p}$ to reconstruct $\overline{\boldsymbol{x}}_{t_2}^{j}$ by inputting ${\bf X}_{t_2}^{[j ; \Delta s]}$:
\begin{equation}
	\begin{split}
		&{\mathcal G}^{p}: 
		{\bf X}_{t_1}^{[j ; \Delta s]} \stackrel{\mathcal{G}_{E}}{\longrightarrow} 
		{\boldsymbol f}_{t_{1}}^{j}
		\stackrel{GEM}{\longrightarrow} 
		\widetilde{{\boldsymbol f}}_{t_{2}}^j 
		\stackrel{{\mathcal G}_{D}}{\longrightarrow} 
		\widetilde{{\boldsymbol x}}_{t_2}^j \\
		&{\mathcal G}^{r}:
		{\bf X}_{t_2}^{[j ; \Delta s]} \stackrel{\mathcal{G}_{E}}{\longrightarrow} 
		{\boldsymbol f}_{t_{2}}^{j} 
		\stackrel{\mathcal{G}_{D}}{\longrightarrow} 
		\overline{{\boldsymbol x}}_{t_2}^j
		\label{eq:7}
	\end{split}
\end{equation}
Prediction generator $\bf{\mathcal{G}}^{p}$ and reconstruction generator $\bf{\mathcal{G}}^{r}$ share feature encoder ${\mathcal G}_{E}$ and feature decoder ${\mathcal G}_{D}$.
We use the project head ${\it h} (\cdot)$ \cite{42chen2020simple} to map ${\boldsymbol f}_{t_1}^j$, $\widetilde{\boldsymbol{f}}_{t_2}^j$, ${\boldsymbol f}_{t_2}^j$ to ${\boldsymbol z}_{t_1}^j$, $\widetilde{\boldsymbol{z}}_{t_2}^j$, ${\boldsymbol z}_{t_2}^j$ for evolution reinforcement learning:
\begin{equation}
	\boldsymbol{z} 
	= h (\boldsymbol{f})
	= \boldsymbol{W}^{(2)}
	(ReLU(\boldsymbol{W}^{(1)} \cdot GAP(\boldsymbol{f})))
	\label{eq:8}
\end{equation}
where $\boldsymbol{W}^{(1)}$, $\boldsymbol{W}^{(2)}$ are weight matrixes, $GAP(\cdot)$ indicates global average pooling. We hope predicted $\widetilde{\boldsymbol{z}}_{t_2}^j$ and real ${\boldsymbol z}_{t_2}^j$ should be as similar as possible, while predicted $\widetilde{\boldsymbol{z}}_{t_2}^j$ and real ${\boldsymbol z}_{t_1}^j$ should be different. Thus, we build the evolution reinforcement learning based on contrastive loss:
\begin{equation}
	\mathcal{L}_{ERM} (\mathcal{G}^p,\mathcal{G}^r) = 
	-log \frac{exp(Sim(\widetilde{\boldsymbol{z}}_{t_2}^j,{\boldsymbol z}_{t_2}^j) / \tau)}{exp(Sim(\widetilde{\boldsymbol{z}}_{t_2}^j,{\boldsymbol z}_{t_1}^j) / \tau)}
	\label{eq:9}
\end{equation}
where $\tau$=1 is the temperature factor, $Sim(\cdot)$ is the cosine similarity metric. ERM also further distills the function of the feature encoder and feature decoder.

\subsection{Discriminators of SHENet}
For the predictively generated $\widetilde{\boldsymbol{x}}_{t_2}^j$ and the reconstructed $\overline{\boldsymbol{x}}_{t_2}^j$, we firstly use a quality discriminator $\mathcal{D}^q$ to ensure their image quality is realistic:
\begin{equation}
	\mathcal{D}^q (\widetilde{\boldsymbol{x}}_{t_2}^j) \rightarrow T/F, ~~ 
	\mathcal{D}^q (\overline{\boldsymbol{x}}_{t_2}^j) \rightarrow T/F
	\label{eq:20}
\end{equation}
Furthermore, we consider that pathological manifestation of $\widetilde{\boldsymbol{x}}_{t_2}^j$ and $\overline{\boldsymbol{x}}_{t_2}^j$ should show the response of $\boldsymbol{x}_{t_1}^j$ after anti-VEGF injection. 
In other words, $\widetilde{\boldsymbol{x}}_{t_2}^j$ and $\overline{\boldsymbol{x}}_{t_2}^j$ should be paired with $\boldsymbol{x}_{t_1}^j$. 
Thus we also use a pair discriminator $\mathcal{D}^p$ to make the decision:
\begin{equation}
	\mathcal{D}^p (\widetilde{\boldsymbol{x}}_{t_2}^j, \boldsymbol{x}_{t_1}^j) \rightarrow T/F, ~~
	\mathcal{D}^p (\overline{\boldsymbol{x}}_{t_2}^j, \boldsymbol{x}_{t_1}^j) \rightarrow T/F
	\label{eq:21}
\end{equation}
In SHENet, both $\mathcal{D}^q$ and  $\mathcal{D}^p$ follow the popular PatchGAN \cite{32isola2017image}.

\subsection{Adversarial Training}
In the training process, we use $\mathcal{L}_{L1}$ measures the L1 distance between output images and ground truth:
\begin{equation}
	\begin{aligned}
	& \mathcal{L}_{L1} (\mathcal{G}^p) = \mathbb{E} [|| \boldsymbol{x}_{t_2}^j - \mathcal{G}^p ({\bf X}_{t_1}^{[j ; \Delta s]}) {||}_1] \\
	& \mathcal{L}_{L1} (\mathcal{G}^r) = \mathbb{E} [|| \boldsymbol{x}_{t_2}^j - \mathcal{G}^r ({\bf X}_{t_2}^{[j ; \Delta s]}) {||}_1]
	\label{eq:10}
	\end{aligned}
\end{equation}
A pair discriminator $\mathcal{D}^p$ makes the decision whether ($\widetilde{\boldsymbol{x}}_{t_2}^j$, $\boldsymbol{x}_{t_1}^j$) and ($\overline{\boldsymbol{x}}_{t_2}^j$, $\boldsymbol{x}_{t_1}^j$) are paired:
\begin{equation}
	\begin{aligned}
		& \mathcal{L}_{GANp} (\mathcal{G}^p,\mathcal{D}^p) = 
		\mathbb{E} [log \mathcal{D}^p (\boldsymbol{x}_{t_1}^j,\boldsymbol{x}_{t_2}^j)] 
		+ \mathbb{E} [log ( 1 - \mathcal{D}^p (\boldsymbol{x}_{t_1}^j , \mathcal{G}^p ( {\bf X}_{t_1}^{[j ; \Delta s]})) )] \\
		& \mathcal{L}_{GANr} (\mathcal{G}^r,\mathcal{D}^p) = 
		\mathbb{E} [log \mathcal{D}^p (\boldsymbol{x}_{t_1}^j,\boldsymbol{x}_{t_2}^j)] 
		+ \mathbb{E} [log ( 1 - \mathcal{D}^p (\boldsymbol{x}_{t_1}^j , \mathcal{G}^r ( {\bf X}_{t_2}^{[j ; \Delta s]})) )]
		\label{eq:11}
	\end{aligned}
\end{equation}
A quality discriminator $\mathcal{D}^q$ ensures  $\widetilde{\boldsymbol{x}}_{t_2}^j$ and $\overline{\boldsymbol{x}}_{t_2}^j$ are realistic:
\begin{equation}
	\begin{aligned}
		& \mathcal{L}_{GANq} (\mathcal{G}^p , \mathcal{D}^q) = 
		\mathbb{E} [log \mathcal{D}^q (\boldsymbol{x}_{t_2}^j)] 
		+ \mathbb{E} [log ( 1 - \mathcal{D}^q (\mathcal{G}^p ({\bf X}_{t_1}^{[{j} ; \Delta s]})) )] \\
		& \mathcal{L}_{GANq} (\mathcal{G}^r , \mathcal{D}^q) = 
		\mathbb{E} [log \mathcal{D}^q (\boldsymbol{x}_{t_2}^j)] 
		+ \mathbb{E} [log ( 1 - \mathcal{D}^q (\mathcal{G}^r ({\bf X}_{t_2}^{[{j} ; \Delta s]})) )]
		\label{eq:12}
	\end{aligned}
\end{equation}

We combine all losses together and the complete optimization can be represented as:
\begin{equation}
	\begin{split}
		\mathcal{G}^p = & arg \min_{\mathcal{G}^p,\mathcal{G}^r}  \max_{\mathcal{D}^p,\mathcal{D}^q} [ \mathcal{L}_{GANp} (\mathcal{G}^p,\mathcal{D}^p) + \mathcal{L}_{GANp} (\mathcal{G}^r,\mathcal{D}^p) \\
		& + \mathcal{L}_{GANq} (\mathcal{G}^p,\mathcal{D}^q) + \mathcal{L}_{GANq} (\mathcal{G}^r,\mathcal{D}^q) \\
		& + \lambda (\mathcal{L}_{L1} (\mathcal{G}^p) + \mathcal{L}_{L1} (\mathcal{G}^r)) + 
		\mu \mathcal{L}_{ERM} (\mathcal{G}^p,\mathcal{G}^r) ]
		\label{eq:13}
	\end{split}
\end{equation}

\section{Results}

\subsection{Data Acquisition}
According to clinical research, the therapeutic effect of anti-VEGF injection for nAMD lasts about one month, so nAMD patients treated should be followed up one month later to evaluate the therapeutic response of nAMD. 
Then, ophthalmologists can make the decision on whether further treatment is needed. 
However, to the best of our knowledge, there are no large-scale, public, and annotated SD-OCT datasets that can be acquired for our model validation due to the differences in imaging protocols, privacy problems, and lack of medical integration. 
This is also a common problem in the field of medical image processing.

The experimental data includes 46208 paired SD-OCT images obtained from 383 SD-OCT cubes of 22 nAMD patients.
Only one eye per nAMD patient is included in the dataset, namely 22 eyes are included.
Each patient contains about 17 serial SD-OCT cubes captured at different time points. 
During the SD-OCT imaging performed once a month, ophthalmologists gave anti-VEGF injections for nAMD treatment. 
The time interval between any adjacent time points is about one month, that is to say, the predictive single horizon is fixed to one month. 
These SD-OCT cubes were captured by Cirrus SD-OCT device \textit{Carl Zeiss Meditec, Inc., Dublin, CA}. 
The size of an SD-OCT cube is $1024 \times 512 \times 128$. 
Detailed attribute descriptions are presented in Table \refeq{tab:111}.

\begin{table}[!ht]\footnotesize
	\setlength\tabcolsep{2.5pt}
	\begin{center}
		\caption{Detailed attribute descriptions of materials.}
		\label{tab:111}
		\begin{tabular}{lc}
			\hline
			\noalign{\smallskip}	
			
		    Attributes & Values \\
			
			\noalign{\smallskip}
			\hline
			\noalign{\smallskip}
			
			Cube Number & 383 \\
			Patient Number & 22 \\
			Time Range & Jan 2012-Dec 2015 \\
			Time Interval & 1 month \\
			Patient Age	& Avg: 70 years (range: 45-84 years) \\
			Patient Gender & 17 males, 5 females \\
			Type & nAMD \\
			Volume & Avg: 0.8024 mm$^{3}$ (0.0133-7.1147 mm$^{3}$) \\
			SD-OCT Device & Carl Zeiss Meditec, Inc., Dublin, CA \\
			Cube Size & 1024$\times$512$\times$128 \\
			Trim Size & 2mm$\times$6mm$\times$6mm \\
			Injection & anti-VEGF \\
			
			\noalign{\smallskip}
			\hline				
		\end{tabular}
	\end{center}
\end{table}

\subsection{Evaluation \& Comparison}
For the predictively generated post-therapeutic SD-OCT images, qualitative evaluation using human subjective judgment is the most straightforward and effective way to verify the prediction performance of SHENet. 
Besides, we also use three metrics, peak signal-to-noise ratio (PSNR), structural similarity (SSIM) and learned perceptual image patch similarity (LPIPS), to evaluate the prediction performance quantitatively. These three metrics have been previously used for the evaluation of image generation \cite{43wang2021dynamic}. 
PSNR measures the image quality and higher PSNR means less distortion. SSIM measures the structure similarity and higher SSIM means higher structure similarity. 
LPIPS measures the similarity of deep visual features and lower LPIPS means higher feature similarity.
To make LPIPS keep the consistency with PSNR and SSIM, namely higher value indicates better performance, here we use 1-LPIPLS to replace LPIPS.

To highlight the advantages of SHENet in terms of I2I prediction performance, we choose four GANs-based methods for competitive comparison, namely Pix2Pix \cite{32isola2017image}, Pix2PixHD \cite{44wang2018high}, Fundus2Angio \cite{45kamran2020fundus2angio}, Att2Angio \cite{46kamran2021attention2angiogan}. 
Pix2Pix and Pix2PixHD have become popular baselines and are widely used for image generation. 
Fundus2Angio and Att2Angio are the latest GANs-based methods for modality transformation in the field of medical image processing.

\begin{figure}[!ht]
	\centering
	\includegraphics[width=\textwidth]{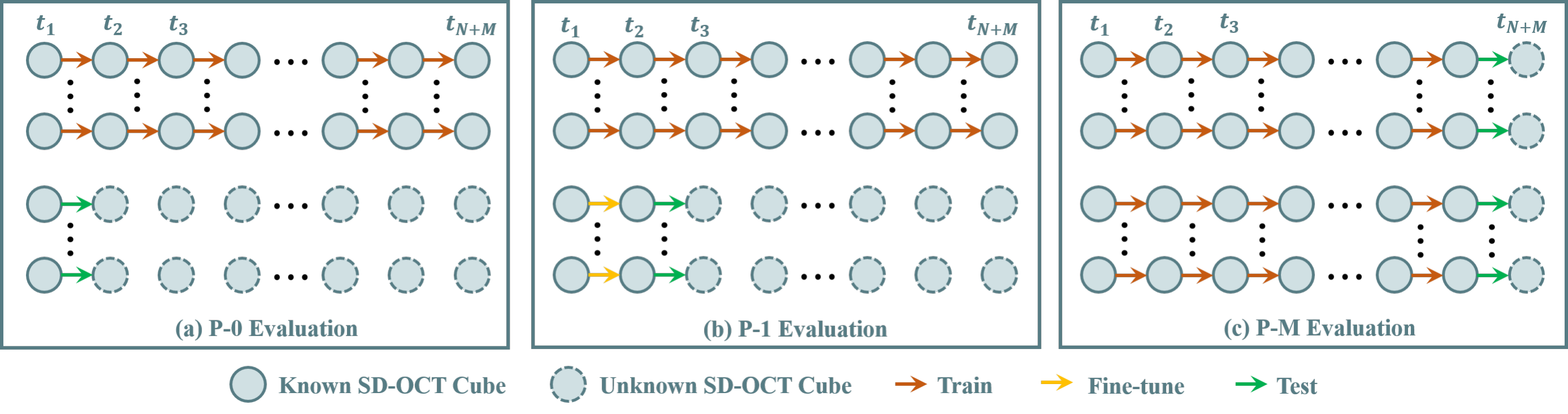}
	\caption{\textbf{Experimental Designs} on P-0 evaluation, P-1 evaluation and P-M evaluation. Each row denotes the serial SD-OCT observations of a nAMD patient and the time interval is one month.}
	\label{fig:7}    % Give a unique label
\end{figure}

\subsection{Experimental Designs}
In order to conform to real clinical application scenes, we design our experiments from three sub-evaluations:

\textbf{P-0 Evaluation.} 
For the new nAMD patients that only have one SD-OCT imaging at time point $t_1$, we want to predict post-therapeutic SD-OCT images at time point $t_2$. 
Thus, the SD-OCT images from all other patients are used for training SHENet. 
We use five-fold cross-validation until all patients are tested, as shown in Fig. \ref{fig:7}(a).

\textbf{P-1 Evaluation.} 
For the nAMD patients that have two SD-OCT imaging at time points $t_1$ and $t_2$, we would like to predict post-therapeutic SD-OCT images at time point $t_3$. 
Thus, we transfer the model parameters from P-0 evaluation and fine-tune SHENet only using $({\bf X}_{t_1},{\bf X}_{t_2})$. 
In the model inference stage, we take ${\bf X}_{t_2}$ as model input to predictively generate ${\bf X}_{t_3}$.
We use five-fold cross-validation until all patients are tested, as shown in Fig. \ref{fig:7}(b).

\textbf{P-M Evaluation.} 
For the nAMD patients that have owned multiple regular SD-OCT imaging, we would like to predict consequent post-therapeutic SD-OCT images. 
Thus, we remain the last two SD-OCT cubes of all patients for model validation and the remaining SD-OCT cubes are used for training model, as shown in Fig. \ref{fig:7}(c).

\subsection{Implementation Details}
The experiment is constructed in a hardware condition with Intel Xeon CPU, one GeForce RTX 3090 GPU and 128 GB RAM, and a software condition with Python3.5 and Pytorch.

For the input multiple B-scans, we choose $\Delta s = 3$ and zero-padding operation is used if $j - \Delta s < 0$ or $j + \Delta s > 128$, thus the model input is a three-channel SD-OCT image.
The number of encoding blocks and decoding blocks is 5.
The output dimensions of 5 encoding blocks are \{128, 256, 512, 1024, 2048\} respectively.
The output feature size of the feature encoder is $16 \times 16 \times 2048$, which means the vertex number of a fully-connected graph in GEM is $16 \times 16$.
In the loss function, $\lambda = 100$ and $\mu = 10$ balance the weights among GAN losses, L1 losses and contrastive loss.
Flip and rotation operations are used here for data augmentation.

Adam optimizer with initial learning rate of 0.0001 and weight decay of 0.1 is chosen for model optimization. 
The batch-size is set to 2.
SHENet was trained for 100 epochs and the training was properly completed.

\begin{figure}[!t]
	\centering
	\includegraphics[width=\textwidth]{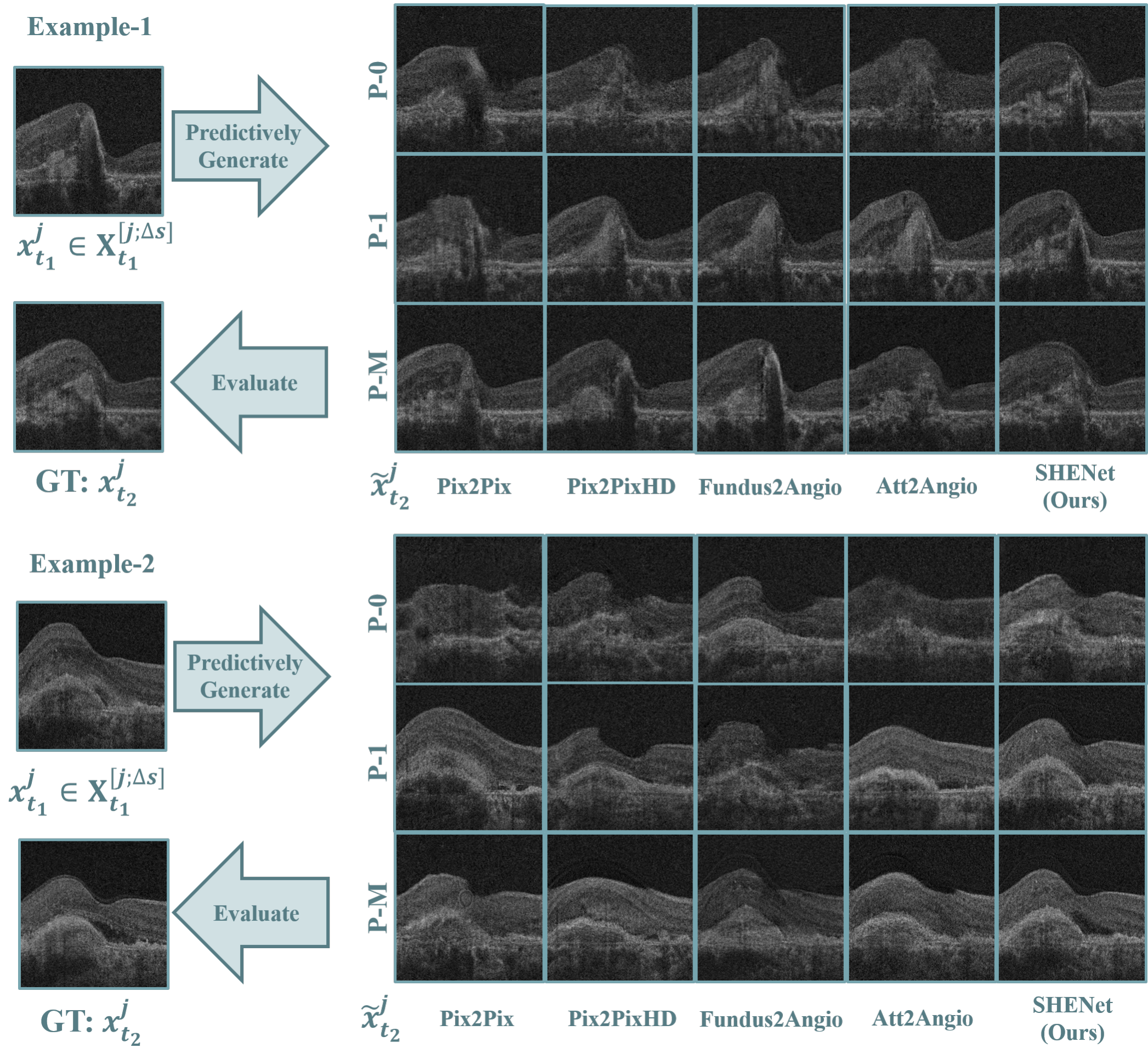}
	\caption{\textbf{Qualitative Comparison} by visualizing the examples from two different nAMD patients. We respectively input a little piece of B-scans $\mathbf{X}_{t_{1}}^{[j ; \Delta s]}$ to Pix2Pix \cite{32isola2017image}, Pix2PixHD \cite{44wang2018high}, Fundus2Angio \cite{45kamran2020fundus2angio}, Att2Angio \cite{46kamran2021attention2angiogan} and the proposed SHENet to predictively generate $\widetilde{\boldsymbol{x}}_{t_2}^j$ based on three experimental designs (P-0, P-1 and P-M), and further evaluate them with ground truth $\boldsymbol{x}_{t_2}^j$.}
	\label{fig:4}    % Give a unique label
\end{figure}

\subsection{Qualitative Evaluation.} 
We qualitatively compare our SHENet with competing methods by visualizing the examples from two different nAMD patients based on P-0, P-1 and P-M evaluations, as shown in Fig. \ref{fig:4}. 
In terms of image quality, benefiting from adversarial training, all methods can produce clear and unblurred SD-OCT images.
Besides, SHENet can stably maintain the structural integrity of the retina to achieve the better visual effects and actually predict the status of nAMD one month later after anti-VEGF injections.

\begin{table}[!ht]\footnotesize
	\setlength\tabcolsep{2pt}
	\begin{center}
		\caption{\textbf{Quantitative Comparison with Other Competing Methods}, based on three experimental designs.}
		\label{tab:1}
		\begin{tabular}{l|ccc|ccc|ccc}
			\hline\noalign{\smallskip}
			
			\multirow{2}{*}{Methods} & \multicolumn{3}{c}{\textbf{P-0}} & \multicolumn{3}{|c|}{\textbf{P-1}} & \multicolumn{3}{c}{\textbf{P-M}} \\
			
			& PSNR & SSIM & 1-LPIPS & PSNR & SSIM & 1-LPIPS & PSNR & SSIM & 1-LPIPS \\
			
			\noalign{\smallskip}
			\hline
			\noalign{\smallskip}
			
			Pix2Pix \cite{32isola2017image}  & 21.114 & 0.273 & 0.552 & 21.272 & 0.276 & 0.559 & 21.481 & 0.284 & 0.568 \\
			Pix2PixHD \cite{44wang2018high} & 21.415 & 0.286 & 0.581 & 21.525 & 0.290 & 0.585 & 21.764 & 0.299 & 0.591 \\
			Fundus2Angio \cite{45kamran2020fundus2angio} & 21.333 & 0.288 & 0.569 & 21.471 & 0.293 & 0.568 & 21.679 & 0.301 & 0.576 \\ 
			Att2Angio \cite{46kamran2021attention2angiogan} & 21.466 & 0.294 & 0.589 & 21.716 & 0.311 & 0.591 & 22.149 & 0.321 & 0.607 \\
			SHENet (Ours) & \pmb{23.659} & \pmb{0.326} & \pmb{0.609} & \pmb{23.875} & \pmb{0.337} & \pmb{0.626} & \pmb{24.198} & \pmb{0.349} & \pmb{0.642} \\
			
			\noalign{\smallskip}
			\hline
		\end{tabular}
	\end{center}
\end{table}

\subsection{Quantitative Evaluation.} 
We quantitatively evaluate the prediction performance based on P-0, P-1 and P-M evaluations using PSNR, SSIM and 1-LPIPS metrics, as shown in Table \ref{tab:1}. 
For P-0 and P-1 evaluations, we calculate the average of five-fold cross-validation as the final results.
Overall, SHENet obtains consistently superior results than competing methods on three experimental designs.
Among three experimental designs, SHENet achieves the best prediction performance on P-M evaluation (PSNR: 24.198, SSIM: 0.349, 1-LPIPS: 0.642), followed by P-1 evaluation (PSNR: 23.875, SSIM: 0.337, 1-LPIPS: 0.626), and P-0 evaluation (PSNR: 23.659, SSIM: 0.326, 1-LPIPS: 0.609) has the worst results, which shows the consistency with qualitative evaluation in Fig. \ref{fig:4}. 
That is because the difference among patients is significantly greater than that among serial SD-OCT observations from the same patient. 
We consider that the gap will narrow by further collecting more samples from more patients. 

\begin{figure}[!ht]
	\centerline{\includegraphics[width=\columnwidth]{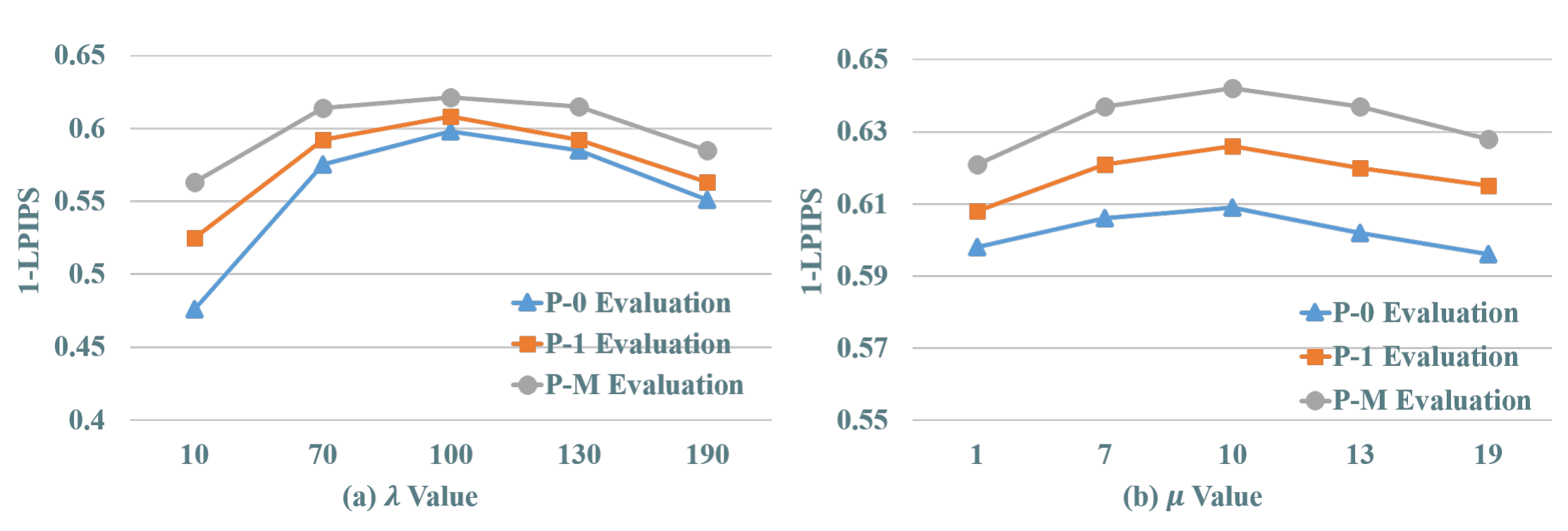}}
	\caption{\textbf{Hyper-parameter Tuning.} Disease evolution prediction 1-LPIPS values with different hyper-parameters $\lambda$ and $\mu$ based on three experimental designs.}
	\label{fig:6}    % Give a unique label
\end{figure}

\subsection{Ablation Analysis}

\textbf{Evaluation of Hyper-parameter Values.}
We analyze the influence of hyper-parameters $\lambda$ and $\mu$ in loss function with the experiment. 
We first fix $\mu$ to be 1 and vary $\lambda$ from 10 to 190 with interval 10. 
As shown in Fig. \ref{fig:6}(a), 1-LPIPS metrics of three evaluations are consistently increasing when rising the of $\lambda$ from 10 to 100 and consistently decreasing when further rising $\lambda$. 
We then fix the value of $\lambda$ to 100 and vary $\mu$ from 1 to 19 with interval 1. 
As shown in Fig. \ref{fig:6}(b), the performance of SHENet achieves the highest values when $\mu = 10$. 

\begin{table}[!ht]\footnotesize
	\setlength\tabcolsep{2.5pt}
	\begin{center}
		\caption{\textbf{Results Comparison with Different Inputs} (single B-scan and multiple B-scans), based on three experimental designs.}
		\label{tab:2}
		\begin{tabular}{clllc}
			\hline\noalign{\smallskip}
			\qquad\qquad\qquad & Model Inputs \qquad\qquad\qquad & PSNR \qquad\qquad & SSIM \qquad\qquad & 1-LPIPS \\
			\noalign{\smallskip}
			\hline
			\noalign{\smallskip}
			\multirow{2}*{\textbf{P-0}} & Single B-scan & 17.542 & 0.242 & 0.513 \\
			~ & Multiple B-scans & \pmb{23.659} & \pmb{0.326} & \pmb{0.609} \\
			\hline
			\noalign{\smallskip}
			\multirow{2}*{\textbf{P-1}} & Single B-scan & 17.567 & 0.244 & 0.515 \\
			~ & Multiple B-scans & \pmb{23.875} & \pmb{0.337} & \pmb{0.626} \\
			\hline
			\noalign{\smallskip}
			\multirow{2}*{\textbf{P-M}} & Single B-scan & 17.585 & 0.250 & 0.521 \\
			~ & Multiple B-scans & \pmb{24.198} & \pmb{0.349} & \pmb{0.642} \\
			\hline		
		\end{tabular}
	\end{center}
\end{table}

\textbf{Evaluation of Model Inputs.}
In terms of model input, we investigate the difference between single B-scan and multiple B-scans, and the quantitative comparisons are recorded in Table \ref{tab:2}. 
We find that the prediction results using multiple B-scans as model input demonstrate significant improvements to those using a single B-scan as model input on three experimental designs.

\begin{table}[!ht]\footnotesize
	\setlength\tabcolsep{2.5pt}
	\begin{center}
		\caption{\textbf{Results Comparison by Stacking GEM, $\mathbf{\mathcal{G}}^r$ and ERM One by One}, based on three experimental designs.}
		\label{tab:3}
		\begin{tabular}{ccc|ccc|ccc|ccc}
			\hline\noalign{\smallskip}
			
			\multirow{2}{*}{GEM} & \multirow{2}{*}{$\mathcal{G}^r$} \qquad & \multirow{2}{*}{ERM} & \multicolumn{3}{c}{\textbf{P-0}} & \multicolumn{3}{|c|}{\textbf{P-1}} & \multicolumn{3}{c}{\textbf{P-M}} \\
			
			& & & PSNR & SSIM & 1-LPIPS & PSNR & SSIM & 1-LPIPS & PSNR & SSIM & 1-LPIPS \\
			
			\noalign{\smallskip}
			\hline\noalign{\smallskip}
			
			$\times$ & $\times$ & $\times$ & 21.188 & 0.276 & 0.559 & 21.277 & 0.281 & 0.563 & 21.485 & 0.288 & 0.574 \\
			$\surd$  & $\times$ & $\times$ & 21.842 & 0.285 & 0.571 & 21.943 & 0.289 & 0.577 & 22.197 & 0.295 & 0.588 \\
			$\surd$  & $\surd$  & $\times$ & 22.648 & 0.292 & 0.588 & 22.791 & 0.298 & 0.594 & 23.016 & 0.309 & 0.611 \\
			$\surd$  & $\surd$  & $\surd$  & \pmb{23.659} & \pmb{0.326} & \pmb{0.609} & \pmb{23.875} & \pmb{0.337} & \pmb{0.626} & \pmb{24.198} & \pmb{0.349} & \pmb{0.642} \\
			\noalign{\smallskip}
			\hline				
		\end{tabular}
	\end{center}
\end{table}

\textbf{Evaluation of Model Architecture.}
In the training process, we build our single-horizon disease evolution in the high-dimensional latent space based on the cooperation of GEM, $\mathcal{G}^r$ and ERM. 
We investigate the impact of three components by stacking them one by one. 
As shown in Table \ref{tab:3}, SHENet achieves better results than others when considering these three components jointly. 
This evidences that SHENet effectively imprisons the process of disease evolution in GEM and further reinforces the process by $\mathcal{G}^r$+ERM. 
Therefore, adopting all three can improve prediction performance. 

\textbf{Evaluation of Discriminator.}
To verify the relevance of two discriminators ($\mathcal{D}^q$, $\mathcal{D}^p$), we separately analyze the prediction results of SHENet when one of them is reserved. 
When only using one of the two, quantitative metrics in Table \ref{tab:4} show performance degradation. 
Besides, we also observe that single $\mathcal{D}^p$ performs better than single $\mathcal{D}^q$, as $\mathcal{D}^p$ can also control the image quality to a certain extent. 
Thus, it shows the necessity to use two independent discriminators to respectively control the image quality and the pathological characterization.

\begin{table}[!ht]\footnotesize
	\setlength\tabcolsep{3pt}
	\begin{center}
		\caption{\textbf{Results Comparison using $\mathbf{\mathcal{D}}^p$, $\mathbf{\mathcal{D}}^q$ or the both}, based on three experimental designs.}
		\label{tab:4}
		\begin{tabular}{cc|ccc|ccc|ccc}
			\hline\noalign{\smallskip}
			
			\multirow{2}{*}{$\mathcal{D}^q$} \qquad & \multirow{2}{*}{$\mathcal{D}^p$} \qquad & \multicolumn{3}{c}{\textbf{P-0}} & \multicolumn{3}{|c|}{\textbf{P-1}} & \multicolumn{3}{c}{\textbf{P-M}} \\
			
			& & PSNR \qquad & SSIM \qquad & 1-LPIPS & PSNR \qquad & SSIM \qquad & 1-LPIPS & PSNR \qquad & SSIM \qquad & 1-LPIPS \\
			
			\noalign{\smallskip}
			\hline
			\noalign{\smallskip}
			
			$\surd$  & $\times$ & 23.574 & 0.320 & 0.602 & 23.783 & 0.325 & 0.618 & 24.088 & 0.337 & 0.633 \\
			$\times$ & $\surd$  & 23.613 & 0.322 & 0.607 & 23.829 & 0.331 & 0.622 & 24.165 & 0.346 & 0.639 \\
			$\surd$  & $\surd$  & \pmb{23.659} & \pmb{0.326} & \pmb{0.609} & \pmb{23.875} & \pmb{0.337} & \pmb{0.626} & \pmb{24.198} & \pmb{0.349} & \pmb{0.642} \\ 
			
			\noalign{\smallskip}
			\hline
		\end{tabular}
	\end{center}
\end{table}

\section{Discussion}
In this paper, according to the actual clinical requirement, we explore the possibility of predictively generating post-therapeutic SD-OCT images based on pre-therapeutic SD-OCT images with nAMD, and propose a single-horizon disease evolution network (SHENet) to solve it. 
SHENet learns the process of disease evolution in the high-dimensional latent space, rather than performing pixel-to-pixel prediction. 
This has the advantage of eliminating the influence of speckle noise and redundant background context on the prediction. 
Considering several inherent characteristics of medical images different from other modal data, we choose single-horizon prediction rather than multi-horizon prediction to simplify the problem.
In other words, we only predictively generate post-therapeutic SD-OCT images with a one-month time interval, and would not continue to predict the nAMD status of the third month.

From Fig. \ref{fig:4}, we can observe structural damage of the retina from competing methods, but SHENet can stably maintain the structural integrity of the retina to achieve a better visual effect.
This benefits from two improvements to the proposed model:
1) SHENet uses two discriminators to respectively manage image quality and pathological characterization, but other competing methods only use one discriminator;
2) SHENet imprisons the process of disease evolution in the high-dimensional latent space, which makes the feature encoder and feature decoder concentrate on the restoration of image details.

\begin{figure}[htb]
	\centering
	\includegraphics[width=0.55\textwidth]{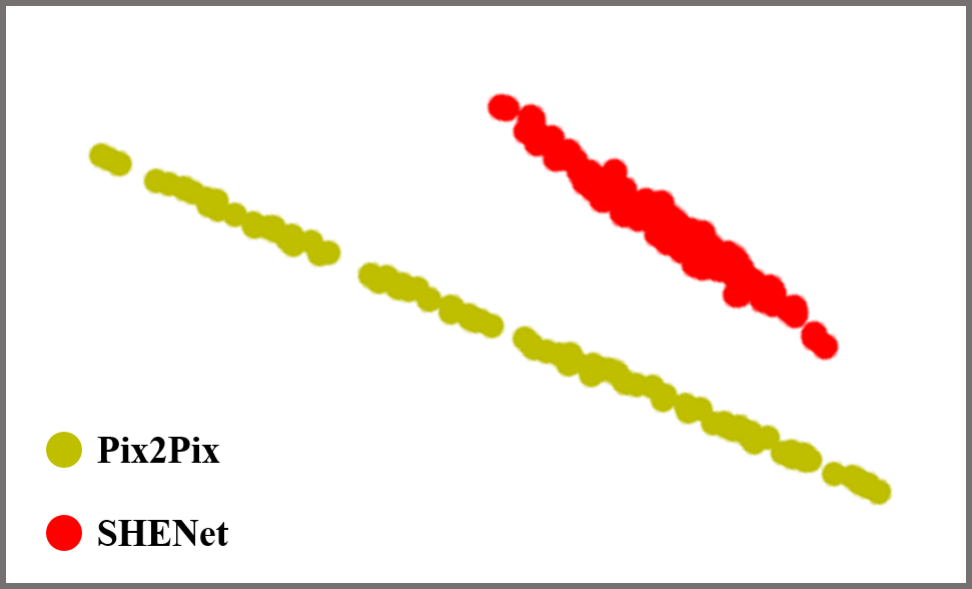}
	\caption{Visualization of latent space of Pix2Pix method and our SHENet from sequential observations.}
	\label{fig:14}    % Give a unique label
\end{figure}

Furthermore, we should pay more attention to the correctness of predictive generation of nAMD, namely whether SHENet can actually predict the status of nAMD one month later after anti-VEGF injection.
Generally speaking, reduction of nAMD volume is the best indicator of therapeutic response, but texture change and improvement of additional diseases also need to be taken into account.
In example-1 in Fig. \ref{fig:4}, by comparing the pre-therapeutic SD-OCT image with the post-therapeutic ground truth, nAMD is significantly reduced in size, which shows an excellent therapeutic response.
In example-2 in Fig. \ref{fig:4}, although the nAMD volume does not change significantly, the thickness of the retina becomes thinner, which still indicates an effective treatment as additional diseases are improved. 
By qualitative comparisons, the predicted visual nAMD of SHENet is closer to ground truth than that of competing methods, demonstrating that SHENet owns a more powerful ability to learn the disease evolution.
We also visualize the latent space of Pix2Pix and our SHENet from sequential observations for further comparison. 
As observed in Fig. \refeq{fig:14}, the image features generated by our SHENet are more high-density than those generated by Pix2Pix. 
That is because our designed GEM and ERM can filter out disease-unrelated information effectively. High-quality image features can provide a good basis for disease evolution learning in the latent space.

\begin{figure}[htb]
	\centering
	\includegraphics[width=\textwidth]{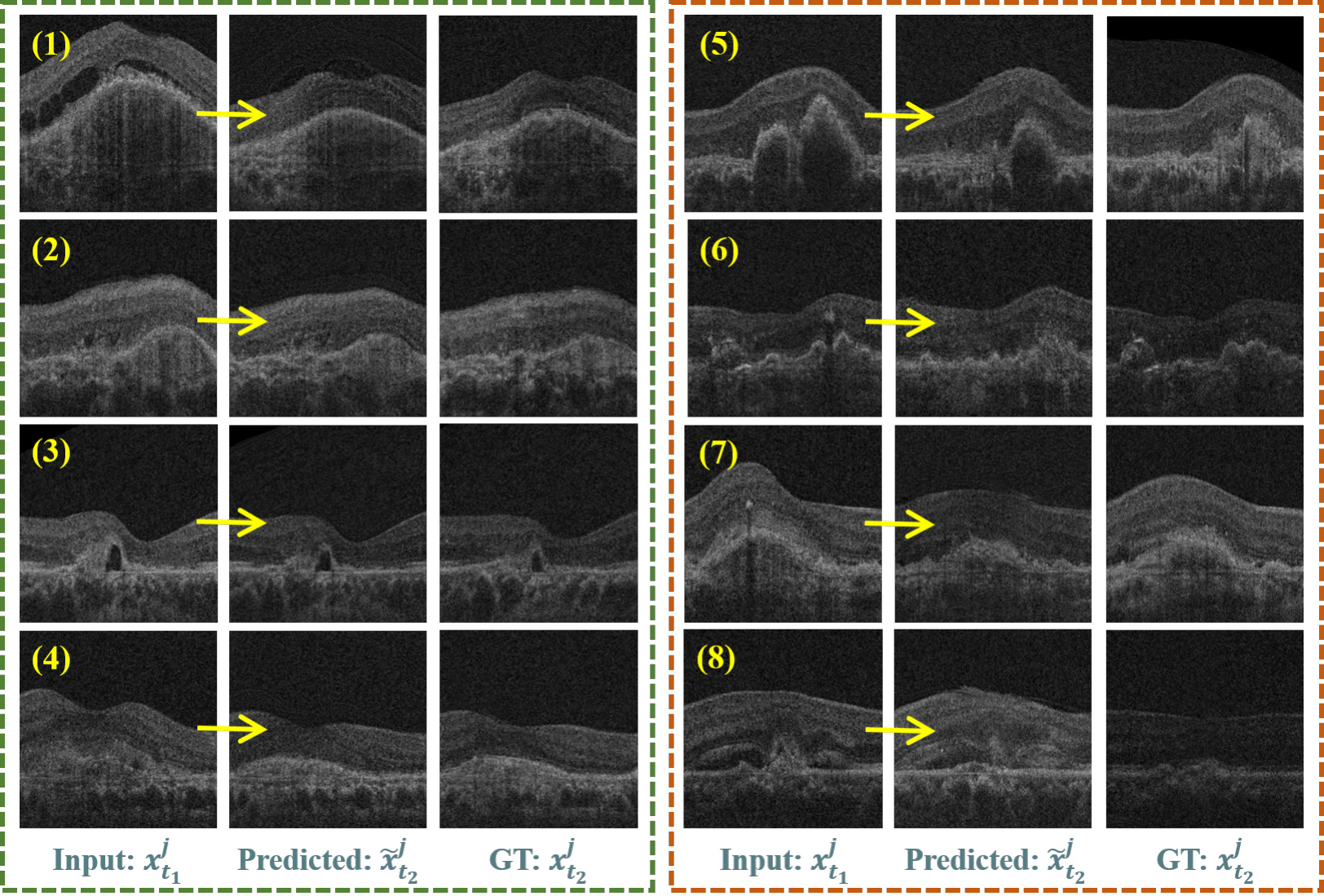}
	\caption{Cases of predicted post-therapeutic SD-OCT images from P-0 evaluation, where (1-4) show the best results and (5-8) show the worst results.}
	\label{fig:15}    % Give a unique label
\end{figure}

From Tables \ref{tab:1}-\ref{tab:4}, we also observe lower PSNR \& SSIM metrics than 1-LPIPS metric, and we consider this is caused by inherent severe speckle noise of SD-OCT images.
Although denoising techniques can reduce the impact of speckle noise, the image quality would also become more blurry and the image details would be lost. 
Considering that the feature encoder owns the denoising ability, we do not apply any denoising technique during the pre-processing, avoiding an overly prediction model. 
In SHENet, to make the predicted images look realistic, the feature decoder adds random speckle noise to the generated SD-OCT images. 
PSNR \& SSIM metrics are evaluated based on the whole SD-OCT images, thus speckle noise can result in low quantitative values, but 1-LPIPS metric is evaluated based on the deep visual features that show higher values than PSNR \& SSIM.

To explore the influence of the number of historical observations on predictive results, we conducted P-0 evaluation, P-1 evaluation and P-M evaluation in our experiments. 
These three evaluations used zero, one and multiple historical observations respectively from the testing patients for model training. 
From Table \ref{tab:1}, we can find that all methods achieve the best results on P-M evaluation and have the worst results on P-0 evaluation, demonstrating that more historical observations of the same patients can improve their own prediction performance. 
Although our dataset contains only 22 nAMD patients, each patient contains about 17 sequential SD-OCT cubes captured at different time points with regular medicine injections. 
A total of 46208 SD-OCT image pairs are used for our model validation. 
Besides, common data augmentation operations (rotation and horizontal flipping) are used. 
Therefore, our method does not occur the overfitting phenomenon, which could be observed from the results of P-0 evaluation. 
In the future, we will validate our method on other medical images, such as fundus photos for AMD and OCT for DME.

Fig. \ref{fig:15} shows more cases of predicted post-therapeutic SD-OCT images from P-0 evaluation, where green dotted box denotes the best results and crimson dotted box denotes the worst results.
Overall, SHENet has the ability to predict the change trend of nAMD status with high image quality, but several flaws still appear.
First, SHENet fails to predict the dramatic texture change of nAMD precisely, as shown in Fig. \ref{fig:15}(5).
Second, due to individual differences, SHENet is hard to model the personalized change rate of nAMD after medicine injection.
For example, SHENet overestimates the effect of medicine injection in Figs. \ref{fig:15}(6-7) and underestimates the effect of medicine injection in Fig. \ref{fig:15}(8).
We consider the main reason is the limited number of patients, leading to SHENet cannot learn comprehensive information due to the complexity of nAMD.
We believe that SHENet can be improved significantly after learning from more nAMD patients, and we will also continually validate SHENet by collecting more data in the future.

Although SHENet can produce high-quality SD-OCT images to visually reflect the status of nAMD one month later after anti-VEGF injection, it also has limitations. 
First, SD-OCT cubes at all time points need to be aligned by manual or automated alignment methods, since SHENet cannot learn the random deviations caused by man-made operations.
Second, the time interval between the model input and ground truth must be consistent in the training process.
Thus, SHENet can only predict the nAMD status after a single horizon and is incapable to predict longer horizons.
Third, each time point for model training must ensure the same treatment intervention.
SHENet can also be extended to train on the serial SD-OCT cubes without treatment intervention for
predicting the single-horizon progression of nAMD. 

\section{Acknowledgment}
This work was supported in part by the National Natural Science Foundation of China (62172223, 61671242), and the Fundamental Research Funds for the Central Universities (30921013105).


\bibliographystyle{elsarticle-num}
\bibliography{example_paper}

\begin{thebibliography}{10}
\expandafter\ifx\csname url\endcsname\relax
  \def\url#1{\texttt{#1}}\fi
\expandafter\ifx\csname urlprefix\endcsname\relax\def\urlprefix{URL }\fi
\expandafter\ifx\csname href\endcsname\relax
  \def\href#1#2{#2} \def\path#1{#1}\fi

\bibitem{01foss2022development}
A.~Foss, T.~Rotsos, T.~Empeslidis, V.~Chong, Development of macular atrophy in
  patients with wet age-related macular degeneration receiving anti-vegf
  treatment, Ophthalmologica 245~(3) (2022) 204--217.

\bibitem{02tadayoni2021brolucizumab}
R.~Tadayoni, L.~Sararols, G.~Weissgerber, R.~Verma, A.~Clemens, F.~G. Holz,
  Brolucizumab: a newly developed anti-vegf molecule for the treatment of
  neovascular age-related macular degeneration, Ophthalmologica 244~(2) (2021)
  93--101.

\bibitem{03mettu2021incomplete}
P.~S. Mettu, M.~J. Allingham, S.~W. Cousins, Incomplete response to anti-vegf
  therapy in neovascular amd: Exploring disease mechanisms and therapeutic
  opportunities, Progress in Retinal and Eye Research 82 (2021) 100906.

\bibitem{04maguire2016five}
M.~G. Maguire, D.~F. Martin, G.-s. Ying, G.~J. Jaffe, E.~Daniel, J.~E.
  Grunwald, C.~A. Toth, F.~L. Ferris~III, S.~L. Fine, C.~of~Age-related Macular
  Degeneration Treatments Trials (CATT) Research~Group, et~al., Five-year
  outcomes with anti--vascular endothelial growth factor treatment of
  neovascular age-related macular degeneration: the comparison of age-related
  macular degeneration treatments trials, Ophthalmology 123~(8) (2016)
  1751--1761.

\bibitem{05lan2021design}
G.~Lan, J.~Xu, Z.~Hu, Y.~Huang, Y.~Wei, X.~Yuan, H.~Liu, J.~Qin, Y.~Wang,
  Q.~Shi, et~al., Design of 1300 nm spectral domain optical coherence
  tomography angiography system for iris microvascular imaging, Journal of
  Physics D: Applied Physics 54~(26) (2021) 264002.

\bibitem{07gharbiya2018comparison}
M.~Gharbiya, R.~Giustolisi, J.~Marchiori, A.~Bruscolini, F.~Mallone, V.~Fameli,
  M.~Nebbioso, S.~Abdolrahimzadeh, Comparison of short-term choroidal thickness
  and retinal morphological changes after intravitreal anti-vegf therapy with
  ranibizumab or aflibercept in treatment-naive eyes, Current eye research
  43~(3) (2018) 391--396.

\bibitem{08saito2017efficacy}
M.~Saito, M.~Kano, K.~Itagaki, T.~Sekiryu, Efficacy of intravitreal aflibercept
  in japanese patients with exudative age-related macular degeneration,
  Japanese journal of ophthalmology 61~(1) (2017) 74--83.

\bibitem{09yim2020predicting}
J.~Yim, R.~Chopra, T.~Spitz, J.~Winkens, A.~Obika, C.~Kelly, H.~Askham,
  M.~Lukic, J.~Huemer, K.~Fasler, et~al., Predicting conversion to wet
  age-related macular degeneration using deep learning, Nature Medicine 26~(6)
  (2020) 892--899.

\bibitem{10ajana2021predicting}
S.~Ajana, A.~Cougnard-Gr{\'e}goire, J.~M. Colijn, B.~M. Merle, T.~Verzijden,
  P.~T. de~Jong, A.~Hofman, J.~R. Vingerling, B.~P. Hejblum, J.-F. Korobelnik,
  et~al., Predicting progression to advanced age-related macular degeneration
  from clinical, genetic, and lifestyle factors using machine learning,
  Ophthalmology 128~(4) (2021) 587--597.

\bibitem{11banerjee2020prediction}
I.~Banerjee, L.~de~Sisternes, J.~A. Hallak, T.~Leng, A.~Osborne, P.~J.
  Rosenfeld, G.~Gregori, M.~Durbin, D.~Rubin, Prediction of age-related macular
  degeneration disease using a sequential deep learning approach on
  longitudinal sd-oct imaging biomarkers, Scientific reports 10~(1) (2020)
  1--16.

\bibitem{12yan2021genome}
Q.~Yan, Y.~Jiang, H.~Huang, A.~Swaroop, E.~Y. Chew, D.~E. Weeks, W.~Chen,
  Y.~Ding, Genome-wide association studies-based machine learning for
  prediction of age-related macular degeneration risk, Translational vision
  science \& technology 10~(2) (2021) 29--29.

\bibitem{13&17bhuiyan2020artificial}
A.~Bhuiyan, T.~Y. Wong, D.~S.~W. Ting, A.~Govindaiah, E.~H. Souied, R.~T.
  Smith, Artificial intelligence to stratify severity of age-related macular
  degeneration (amd) and predict risk of progression to late amd, Translational
  vision science \& technology 9~(2) (2020) 25--25.

\bibitem{14schmidt2018machine}
U.~Schmidt-Erfurth, H.~Bogunovic, A.~Sadeghipour, T.~Schlegl, G.~Langs, B.~S.
  Gerendas, A.~Osborne, S.~M. Waldstein, Machine learning to analyze the
  prognostic value of current imaging biomarkers in neovascular age-related
  macular degeneration, Ophthalmology Retina 2~(1) (2018) 24--30.

\bibitem{15rohm2018predicting}
M.~Rohm, V.~Tresp, M.~M{\"u}ller, C.~Kern, I.~Manakov, M.~Weiss, D.~A. Sim,
  S.~Priglinger, P.~A. Keane, K.~Kortuem, Predicting visual acuity by using
  machine learning in patients treated for neovascular age-related macular
  degeneration, Ophthalmology 125~(7) (2018) 1028--1036.

\bibitem{16diack2021baseline}
C.~Diack, D.~Schwab, V.~Cosson, V.~Buchheit, N.~Mazer, N.~Frey, A baseline
  score to predict response to ranibizumab treatment in neovascular age-related
  macular degeneration, Translational Vision Science \& Technology 10~(6)
  (2021) 11--11.

\bibitem{18rossant2021normalization}
F.~Rossant, M.~Paques, Normalization of series of fundus images to monitor the
  geographic atrophy growth in dry age-related macular degeneration, Computer
  Methods and Programs in Biomedicine 208 (2021) 106234.

\bibitem{19zhang2021integrated}
Y.~Zhang, X.~Zhang, Z.~Ji, S.~Niu, T.~Leng, D.~L. Rubin, S.~Yuan, Q.~Chen, An
  integrated time adaptive geographic atrophy prediction model for sd-oct
  images, Medical Image Analysis 68 (2021) 101893.

\bibitem{20reiter2020investigating}
G.~S. Reiter, R.~Told, L.~Baumann, S.~Sacu, U.~Schmidt-Erfurth, A.~Pollreisz,
  Investigating a growth prediction model in advanced age-related macular
  degeneration with solitary geographic atrophy using quantitative
  autofluorescence, Retina 40~(9) (2020) 1657--1664.

\bibitem{21nattagh2020oct}
K.~Nattagh, H.~Zhou, N.~Rinella, Q.~Zhang, Y.~Dai, K.~G. Foote, C.~Keiner,
  M.~Deiner, J.~L. Duncan, T.~C. Porco, et~al., Oct angiography to predict
  geographic atrophy progression using choriocapillaris flow void as a
  biomarker, Translational vision science \& technology 9~(7) (2020) 6--6.

\bibitem{22zhang2019multi}
Y.~Zhang, Z.~Ji, S.~Niu, T.~Leng, D.~L. Rubin, Q.~Chen, A multi-scale deep
  convolutional neural network for joint segmentation and prediction of
  geographic atrophy in sd-oct images, in: 2019 IEEE 16th International
  Symposium on Biomedical Imaging (ISBI 2019), IEEE, 2019, pp. 565--568.

\bibitem{23yang2021multi}
Q.~Yang, N.~Anegondi, V.~Steffen, C.~Rabe, D.~Ferrara, S.~S. Gao, Multi-modal
  geographic atrophy lesion growth rate prediction using deep learning,
  Investigative Ophthalmology \& Visual Science 62~(8) (2021) 235--235.

\bibitem{24bogunovic2017prediction}
H.~Bogunovi{\'c}, S.~M. Waldstein, T.~Schlegl, G.~Langs, A.~Sadeghipour,
  X.~Liu, B.~S. Gerendas, A.~Osborne, U.~Schmidt-Erfurth, Prediction of
  anti-vegf treatment requirements in neovascular amd using a machine learning
  approach, Investigative ophthalmology \& visual science 58~(7) (2017)
  3240--3248.

\bibitem{25liu2020prediction}
Y.~Liu, J.~Yang, Y.~Zhou, W.~Wang, J.~Zhao, W.~Yu, D.~Zhang, D.~Ding, X.~Li,
  Y.~Chen, Prediction of oct images of short-term response to anti-vegf
  treatment for neovascular age-related macular degeneration using generative
  adversarial network, British Journal of Ophthalmology 104~(12) (2020)
  1735--1740.

\bibitem{47lee2021post}
H.~Lee, S.~Kim, M.~A. Kim, H.~Chung, H.~C. Kim, Post-treatment prediction of
  optical coherence tomography using a conditional generative adversarial
  network in age-related macular degeneration, Retina 41~(3) (2021) 572--580.

\bibitem{26forshaw2020full}
T.~R.~J. Forshaw, H.~J. Ahmed, T.~W. Kj{\ae}r, S.~Andr{\'e}asson, T.~L.
  S{\o}rensen, Full-field electroretinography in age-related macular
  degeneration: can retinal electrophysiology predict the subjective visual
  outcome of cataract surgery?, Acta ophthalmologica 98~(7) (2020) 693--700.

\bibitem{48pham2022generating}
Q.~T. Pham, S.~Ahn, J.~Shin, S.~J. Song, Generating future fundus images for
  early age-related macular degeneration based on generative adversarial
  networks, Computer Methods and Programs in Biomedicine 216 (2022) 106648.

\bibitem{27goodfellow2014generative}
I.~Goodfellow, J.~Pouget-Abadie, M.~Mirza, B.~Xu, D.~Warde-Farley, S.~Ozair,
  A.~Courville, Y.~Bengio, Generative adversarial nets, Advances in neural
  information processing systems 27 (2014).

\bibitem{28johnson2016perceptual}
J.~Johnson, A.~Alahi, L.~Fei-Fei, Perceptual losses for real-time style
  transfer and super-resolution, in: European conference on computer vision,
  Springer, 2016, pp. 694--711.

\bibitem{29arjovsky2017wasserstein}
M.~Arjovsky, S.~Chintala, L.~Bottou, Wasserstein generative adversarial
  networks, in: International conference on machine learning, PMLR, 2017, pp.
  214--223.

\bibitem{30nowozin2016f}
S.~Nowozin, B.~Cseke, R.~Tomioka, f-gan: Training generative neural samplers
  using variational divergence minimization, in: Proceedings of the 30th
  International Conference on Neural Information Processing Systems, 2016, pp.
  271--279.

\bibitem{31mirza2014conditional}
M.~Mirza, S.~Osindero, Conditional generative adversarial nets, arXiv preprint
  arXiv:1411.1784 (2014).

\bibitem{51yoo2020generative}
T.~K. Yoo, J.~Y. Choi, H.~K. Kim, A generative adversarial network approach to
  predicting postoperative appearance after orbital decompression surgery for
  thyroid eye disease, Computers in biology and medicine 118 (2020) 103628.

\bibitem{49qiu2022improved}
D.~Qiu, Y.~Cheng, X.~Wang, Improved generative adversarial network for retinal
  image super-resolution, Computer Methods and Programs in Biomedicine 225
  (2022) 106995.

\bibitem{50zhang2022bpgan}
J.~Zhang, X.~He, L.~Qing, F.~Gao, B.~Wang, Bpgan: Brain pet synthesis from mri
  using generative adversarial network for multi-modal alzheimer’s disease
  diagnosis, Computer Methods and Programs in Biomedicine 217 (2022) 106676.

\bibitem{35schonfeld2020u}
E.~Schonfeld, B.~Schiele, A.~Khoreva, A u-net based discriminator for
  generative adversarial networks, in: Proceedings of the IEEE/CVF Conference
  on Computer Vision and Pattern Recognition, 2020, pp. 8207--8216.

\bibitem{36park2019semantic}
T.~Park, M.-Y. Liu, T.-C. Wang, J.-Y. Zhu, Semantic image synthesis with
  spatially-adaptive normalization, in: Proceedings of the IEEE/CVF Conference
  on Computer Vision and Pattern Recognition, 2019, pp. 2337--2346.

\bibitem{37zhang2021robust}
Y.~Zhang, M.~Li, S.~Yuan, Q.~Liu, Q.~Chen, Robust region encoding and layer
  attribute protection for the segmentation of retina with multifarious
  abnormalities, Medical Physics 48~(12) (2021) 7773--7789.

\bibitem{32isola2017image}
P.~Isola, J.-Y. Zhu, T.~Zhou, A.~A. Efros, Image-to-image translation with
  conditional adversarial networks, in: Proceedings of the IEEE conference on
  computer vision and pattern recognition, 2017, pp. 1125--1134.

\bibitem{39hu2018squeeze}
J.~Hu, L.~Shen, G.~Sun, Squeeze-and-excitation networks, in: Proceedings of the
  IEEE conference on computer vision and pattern recognition, 2018, pp.
  7132--7141.

\bibitem{40misra2019mish}
D.~Misra, Mish: A self regularized non-monotonic neural activation function,
  arXiv preprint arXiv:1908.08681 4 (2019) 2.

\bibitem{41velivckovic2017graph}
P.~Veli{\v{c}}kovi{\'c}, G.~Cucurull, A.~Casanova, A.~Romero, P.~Lio,
  Y.~Bengio, Graph attention networks, arXiv preprint arXiv:1710.10903 (2017).

\bibitem{42chen2020simple}
T.~Chen, S.~Kornblith, M.~Norouzi, G.~Hinton, A simple framework for
  contrastive learning of visual representations, in: International conference
  on machine learning, PMLR, 2020, pp. 1597--1607.

\bibitem{43wang2021dynamic}
N.~Wang, Y.~Zhang, L.~Zhang, Dynamic selection network for image inpainting,
  IEEE Transactions on Image Processing 30 (2021) 1784--1798.

\bibitem{44wang2018high}
T.-C. Wang, M.-Y. Liu, J.-Y. Zhu, A.~Tao, J.~Kautz, B.~Catanzaro,
  High-resolution image synthesis and semantic manipulation with conditional
  gans, in: Proceedings of the IEEE conference on computer vision and pattern
  recognition, 2018, pp. 8798--8807.

\bibitem{45kamran2020fundus2angio}
S.~A. Kamran, K.~F. Hossain, A.~Tavakkoli, S.~Zuckerbrod, S.~A. Baker, K.~M.
  Sanders, Fundus2angio: A conditional gan architecture for generating
  fluorescein angiography images from retinal fundus photography, in:
  International Symposium on Visual Computing, Springer, 2020, pp. 125--138.

\bibitem{46kamran2021attention2angiogan}
S.~A. Kamran, K.~F. Hossain, A.~Tavakkoli, S.~L. Zuckerbrod,
  Attention2angiogan: Synthesizing fluorescein angiography from retinal fundus
  images using generative adversarial networks, in: 2020 25th International
  Conference on Pattern Recognition (ICPR), IEEE, 2021, pp. 9122--9129.

\end{thebibliography}


\end{document}